# ACCELERATING COSMOLOGIES WITH EXTENDED PRODUCT SPACES

By

**CH'NG HAN SIONG**

**Thesis Submitted to the School of Graduate Studies, Universiti Putra Malaysia, in Fulfilment of the Requirement for the Degree of Master of Science**

**November 2006**

Abstract of thesis presented to the Senate of Universiti Putra Malaysia in fulfilment of the requirement for the degree of Master of Science

# ACCELERATING COSMOLOGIES WITH EXTENDED PRODUCT SPACES

By

**CH'NG HAN SIONG**

**November 2006**

**Chairman:** **Associate Professor Hishamuddin Zainuddin, PhD**

**Institute:** **Advanced Technology**

Accelerating cosmologies in extra dimensional spaces have been studied. These extra dimensional spaces are products of many spaces. The physical behaviors of accelerating cosmologies are investigated from Einstein's field equation in higher dimensional Friedmann-Robertson-Walker (FRW) universe and superstring/M theory points of view. It is found that if some assumptions of flatness are made for sector of the FRW universe, the remaining sector needs to be hyperbolic. These properties are in parallel with those found in the model of superstring/M theory. The extended product made for the superstring model did not show any more new features other than those already found. A similar accelerating phase of this product space cosmology was found with difference in numerical values of the accelerating period.

# ACKNOWLEDGEMENTS

I would like to thank my supervisor, Assoc. Prof. Dr. Hishamuddin Zainuddin, for his introduction of the field of accelerating cosmologies to me. I am very grateful for the freedom he has given me in doing the research. He is undoubtedly a wise researcher knows well the ways to conduct open and sincere discussions with me, and he master well the skills in fine tuning guidance from himself and opinions from me. I also truly appreciate with his positive assessment about my research ability greatly strengthens my endurance to face challenges in doing this research. I would also like to thank Dr. Zainul Abidin Hassan, who has been my co-supervisor for this work.

I am also grateful to the Institute of Advance Technology and Theoretical Studies Laboratory, where this work was initiated. This research was financially supported by the Ministry of Science and Innovation under Pasca Siswazah Scheme.

I certify that an Examination committee has met on 28[th] November 2006 to conduct the final examination of Ch'ng Han Siong on his Master of Science thesis entitled "Accelerating Cosmologies with Extended Product Spaces" in accordance with Universiti Putra Malaysia (Higher Degree) Act 1980 and Universiti Putra Malaysia (Higher Degree) Regulations 1981. The committee recommends that the candidate be awarded the relevant degree. Members of the Examination Committee are as follows:

**MOHD. YUSOF SULAIMAN, PhD**
Professor
Faculty of Science
Universiti Putra Malaysia
(Chairman)

**JUMIAH HASSAN, PhD**
Associate Professor
Faculty of Science
Universiti Putra Malaysia
(Internal Examiner)

**ADEM KILICMAN, PhD**
Associate Professor
Faculty of Science
Universiti Putra Malaysia
(Internal Examiner)

**ROSY THE CHOOI GIM, PhD**
Associate Professor
Faculty of Science
Universiti Sains Malaysia
(External Examiner)

________________________________
**HASANAH MOHD. GHAZALI, PhD**
Professor/Deputy Dean
School of Graduate Studies
Universiti Putra Malaysia

Date:

This thesis submitted to the Senate of Universiti Putra Malaysia and has been accepted as fulfilment of the requirement for the degree of Master of Science. The members of the Supervisory Committee are as follows:

**Hishamuddin Zainuddin, PhD**
Associate Professor
Institute of Advanced Technology
Universiti Putra Malaysia
(Chairman)

**Zainul Abidin Hassan, PhD**
Lecturer
Faculty of Science
Universiti Putra Malaysia
(Member)

____________________
**AINI IDERIS, PhD**
Professor/Dean
School of Graduate Studies
Universiti Putra Malaysia

Date: 8 MARCH 2007

# TABLE OF CONTENTS

# LIST OF NOTATIONS

| | |
|---|---|
| $R(t)$ | Scale factor (if it is function of t) |
| $R$ | Ricci scalar (if it is not function of t) |
| $R_{\mu\nu}$ | Ricci tensor |
| $R^{\sigma}_{\mu\nu\lambda}$ | Riemann tensor |
| $T_{\mu\nu}$ | Energy-momentum tensor |
| $g_{\mu\nu}$ | Metric tensor |
| $G$ | Newton's gravitational constant |
| $\bar{G}$ | Higher dimensional Newton's gravitational constant |
| $\Lambda$ | Cosmological constant |
| $G_{\mu\nu}$ | Einstein tensor |
| $k_i$ | Curvature of i-th sector of product space |
| $m_i$ | Number of dimensions of the i-th sector of product space |
| $R_i$ | Scale factor of i-th sector of product space |
| $\varepsilon$ | Mass-energy density |
| $p_i$ | Pressure of i-th sector of product space |
| $\theta_1^{(i)}, \theta_2^{(i)}, \theta_3^{(i)}, \dots$ | Coordinate of spherical coordinate system for i-th sector of product space |
| $R_{\mu\nu}^{(i)}$ | Ricci tensor for i-th sector of product space |
| $g_{\mu\nu}^{(i)}$ | Metric tensor for i-th sector of product space |

# CHAPTER 1

# INTRODUCTION

## 1.1 History of Accelerating Cosmologies

The Wilkinson Microwave Anisotropy Probe (WMAP) data and the observations of the dimming of type Ia supernovae in distant galaxies showed that the universe is undergoing accelerated expansion at the present epoch [1,2]. Although it is not difficult to find cosmological models that exhibit these features, but usually one will hope that such model can be derivable from a fundamental theory that incorporates both gravity and the standard model of particle physics [3]. Most such efforts are from superstring or M theory framework [3,4,5,6,7,8]. In superstring theory or M theory, one needs a compactification of six or seven internal spaces. There is however no solution for accelerating universe if the internal space is time-independent [4]. Besides this, Einstein's equations in higher dimensions are of interest as generalizing known solutions. There are some of these earlier works that were done in Ref. [9,10,11,12], which considered the extra dimensions in single space. Besides this, extra dimensional space was treated as a source for generating acceleration of the universe.

The no-go theorem is due to the strong energy condition not being violated by either eleven dimensional supergravity or any of the ten dimensional supergravity theories. If the higher dimensional stress tensor satisfies the strong energy condition, so will the lower dimensional stress tensor. But this is not the case in Friedmann-Lemaître-Robertson-Walker (FLRW) (homogeneous and isotropic) universe, because it needs a violation of the strong-energy condition.

The metric of the FLRW form:

$$ds^2 = -d\tau^2 + a(\tau)^2\left(\frac{dr^2}{1-kr^2} + r^2 d\Omega^2\right), \qquad (1.1\text{-}1)$$

where k = -1,0,+1;

$a(\tau)$ = scale factor;

$d\Omega^2 = d\theta^2 + \sin^2\theta d\phi^2$.

The time-time component of 4D Ricci tensor for FLRW is given in [4]

$$R_{00} = -3\frac{\ddot{a}}{a}, \qquad (1.1\text{-}2)$$

where $a$ = scale factor;

$\ddot{a}$ = second order time derivative of the scale factor.

So, in four-dimensional FLRW universe, the universe is accelerating if strong energy condition is violated.

The strong energy condition $R_{00} \geq 0$, is however necessary for lower dimensional stress tensor of supergravity theory. Therefore, we must circumvent this no-go theorem in order to get a viable cosmology from string or M theory. Townsend and N. R. Wohlfarth found that this can be achieved if the condition of time-independence of the internal space is given up [3].

Recently, Townsend and Wohlfarth [3] have worked on a single hyperbolic internal space. Subsequently, this work was generalized by proposing the product of internal spaces. So now, we are interested on product space $R^{3+1} \times R^{m_1} \times R^{m_2} \times H^{m_3}$. These product internal spaces are useful because they provide the situation similar to the hybrid model of inflation [5].

This works can be divided into two parts. For part one, we generalized Einstein's equation to include extra dimensions in the form of product spaces and give some interpretations. Before doing these interpretations, we will make some assumptions for physical behavior of the scale factor. In addition, we assume the universe to be matter-dominated and the spatial curvature for our ordinary three dimensional space is zero. Later, we will turn to consider the accelerating universe in superstring/M-theory point of view. The extra dimensional spaces that will be considered in this context are also product spaces.

In this part two, we need to find a scale factor for the evolution of the universe and from the scale factor, there is a time interval in which the universe is undergoing accelerated expansion.

The objective of this work is to determine the spatial curvature of extra dimensions that is responsible for obtaining accelerated expanding universe. These extra dimensions can be in the form of product spaces.

## 1.2 No-go Theorem for String Theory

The Einstein equation can produce the two following equations [4]:

Raychaudhuri equation: $\dfrac{\ddot{a}}{a} = -\dfrac{4\pi G}{3}(\rho + 3p)$ ; (1.2-1)

Friedmann equations : $\left(\dfrac{\dot{a}}{a}\right)^2 + \dfrac{k}{a^2} = \dfrac{8\pi G \rho}{3}$ ; (1.2-2)

where $\rho$ = energy density,

$p$ = pressure.

For an accelerating cosmology, $\dfrac{\ddot{a}}{a}$ must be positive. Therefore, $\rho + 3p$ must be negative. This violates the strong energy condition [13,7].

From Ref. [4], $\rho + 3p = 2\left(T_{00} - \dfrac{1}{2}g_{00}T^{\lambda}_{\lambda}\right)$

$$= \dfrac{1}{4\pi G} R_{00} \ . \qquad (1.2\text{-}3)$$

Raychaudhuri equation becomes:

$$\dfrac{\ddot{a}}{a} = -\dfrac{1}{3}R_{00} . \qquad (1.2\text{-}4)$$

If we want to get inflation, $R_{00}$ must be negative. But, as it was already mentioned in section 1.1, strong energy condition is not violated by eleven dimensional supergravities or any ten dimensional supergravity theories. Therefore, if the higher dimensional stress tensor satisfies the strong energy condition, then so will the lower dimensional stress tensor.

This implies that [7]

$$R^{(D)}_{00} \geq 0 \text{ only if } R^{(4)}_{00} \geq 0.$$

Actually, this is a consequence from the following deduction [6,7,9]:

We consider a $D$ dimensional metric ansatz:

$$ds_D^2 = A^2(y) ds_4^2(x) + ds_m^2(y), \quad (1.2\text{-}5)$$

where $ds_m^2$ is the metric of some compact, non-singular m-manifold $M$ without boundary with coordinates $y$ and $ds_4^2$ is the metric of the four dimensional space-time with coordinates $x$. $A$ is a smooth, non-vanishing function, known as a warp factor [7,14].

For the compact extra dimensions, we know that [6,7]

$$\left[\int_M A^2\right] R^{(4)}_{00} = \int_M A^2 R^{(D)}_{00}. \quad (1.2\text{-}6)$$

This shows that $R^{(D)}_{00} \geq 0$ only if $R^{(4)}_{00} \geq 0$.

The question now arises on how to circumvent the no-go theorem in the hope of explaining the accelerating cosmology. This solution was given in [3]. There is a time-dependent scalar fields, $\phi$ and a metric $g$, one can consider

$$\tilde{g}_{\mu\nu} = e^{2\phi} g_{\mu\nu} \quad (1.2\text{-}7)$$

as a new, conformally rescaled metric.

If $g_{\mu\nu}$ is the metric of the FLRW cosmology and scalar field, $\phi$ depend only on time, then

$$\tilde{R}_{00} = R_{00} - 3\left[\ddot{\phi}(t) + H(t)\dot{\phi}(t)\right], \qquad (1.2\text{-}8)$$

where $H(t)$ is Hubble 'constant'.

The key point is that positivity of $R_{00}$ does not indicate positivity of $\tilde{R}_{00}$ [3,7].

## 1.3 A Brief of Superstring/ M-Theory

Nowadays, one major problem considered by physicists is that the quantum field theory is incompatible with general relativity. One particular way to reconcile these theories is to modify the quantum field theory in order that it can give rise to the explanation of gravity [15, 16]. This can be done by giving up one of the basic notions of quantum field theory that is the mathematical point description of elementary particles becomes one-dimensional extended objects. The extended objects are called strings. String theory gives a very interesting result that the theory of general relativity is essentially an outcome in this theory. When the distance is very short, the theory can explain very well what is expected, and when the distance is at ordinary range, it yields the explanation as in general relativity. In other words, string theory requires the existence of gravity. In quantum field theory, the point particle forms the world line in a space-time diagram. While for string theory, the corresponding part is called world sheet. The idea is that the fundamental element, strings are of the order of Planck length (about $10^{-35}$ m) and vibrate at resonant frequencies. Every string vibrates at different resonance frequencies and these different resonance frequencies determine the certain types of force. String theory needs 10,11 or 26 space-time dimensions, which is more than the number of dimensions that can be observed. This is called extra dimensions, and the extra dimensions are compactified to a very small size. Physicists also incorporate supersymmetry into the string theory, and then the theory is called superstring theory. There are five different types of superstring theories and they are "Type I", "Type IIA", "Type IIB", "Heterotic-O" and "Heterotic-E". This poses a problem, since one desired property of a unified theory is just a single theory explaining all physical forces. In the 1990s, this problem has however been addressed in the second superstring revolution. These five different superstring theories were found to be different limits of a yet to be known single fundamental theory called M-theory.

# CHAPTER 2

# LITERATURE REVIEW

The need of extra dimensions is common in theories that are unifying gravity and other forces. We could not see these extra dimensions because the scales of these extra dimensions arenot within the present limits of physical detection. Since the unified theory in principle gives all physical forces, scientists thus hope that these theories can explain the phenomena of accelerating universe. In this section, different proposals of the accelerating universe with extra dimensions are reviewed.

## 2.1 Evolution of the Universe with Flat Extra Dimensions

Gu et al. [17] proposed that the evolution of our universe in different eras, including the present accelerating expansion era, is governed not wholly by the matter contents (excepting dark energy), but also by the curvature of the ordinary 3-space and the evolution of extra spaces. They investigate the possibility of this model to accommodate an accelerated expansion and at the same time conforming to observational constraints. They also present several significant features of their model, such as the automatic stabilization of extra dimensions and the explanation to the cosmic coincidence problem (the energy densities of dark energy and dark matter are comparable now).

In their work, they consider a $(3+n+1)$-dimensional space-time where n is the number of extra spatial dimensions. They make an assumption that the three-dimensional ordinary space and the $n$-dimensional extra space are homogeneous and isotropic, and represent this space-time by using two spatial parts of the Robertson-Walker metric as follows:

$$ds^2 = dt^2 - a^2(t)\left(\frac{dr_a^2}{1-k_a r_a^2} + r_a^2 d\Omega_a^2\right) - b^2(t)\left(\frac{dr_b^2}{1-k_b r_b^2} + r_b^2 d\Omega_b^2\right), \qquad (2.1\text{-}1)$$

where $a(t)$ and $b(t)$ are scale factors, and $k_a$ and $k_b$ are curvatures of the ordinary 3-space and the extra space, respectively.

## 2.2 Cosmological Evolution of Homogeneous Universal Extra Dimensions

Bringmann et al. [18] proposed a cosmological model called Homogeneous Universal Extra Dimensions (UED) theory. They present the cosmological solutions of Einstein's field equations for a $(3+n+1)$ dimensional homogeneous and anisotropic universe. They also explain on the interpretation of pressure in higher dimensions and obtain a general relation between pressure and energy density in universal extra dimensions (UED) cosmology. They investigate the solutions with constant extra dimensions during radiation and matter domination era and considered a possible transition between these two states.

For the UED model, the lightest Kaluza-Klein (KK) particle (LKP) could still be present today as a thermal relic, because of KK parity conservation. In addition, it has all the properties of a weakly interacting massive particle (WIMP) if it is neutral and nonbaryonic and is one of the dark matter candidates. The KK photon and the KK neutrino could account for dark matter with $\Omega_M \sim 0.3$, as proposed by the current cosmological concordance model [10], for which one takes the compactification scale of about $R \sim 1 TeV^{-1}$ size.

Their work is to study the evolution of the universe as modeled by a suitable extension of the usual Friedmann equations that is the corresponding Friedmann equations for higher dimensional universe. They concern on solutions with constant or slowly varying extra dimensions without any stabilization mechanism. Despite this, their work is to investigate whether the extra dimensions are varying with time or otherwise. From this perspective, they also did a numerical study on the transition regime between the radiation dominated and matter dominated eras.

### 2.3 Problems with Time-Varying Extra Dimensions or "Cardassian Expansion" as Alternatives to Dark Energy

Cline and Vinet have discussed in Ref. [19] on the mechanisms where the acceleration is due to the presence of extra dimensions, which is proposed in Ref. [12] and Ref. [20]. In [12], the main role is played by a new component, which is some kind of bulk stress energy. This kind of new component could change the form of the Friedmann equations at later times. While in Ref. [20], the acceleration is due to the time-variation of the size of the extra dimensions. Then, they look for accelerated expansion universe by applying these ideas and at the same time, complying with well known physical constraints, which are on the time variation of the four dimensional Newton's gravitational constant or on the possible equation of state of the late form of stress energy.

They tested the idea in [20] against experimental and observational constraints on the constancy over time of the gravitational force, indicating that it goes into a problem. They also examine the relationship between this model and Brans-Dicke theory. They also showed some problems can be overcome by assuming that the extra dimensions are not isotropic. They also show that the Cardassian acceleration model does not obey the weak energy condition in the bulk.

## 2.4 Accelerating Cosmologies from Compactification

Townsend and Wohlfarth [3] have shown that if we give up the condition of time-independence of the internal space, the no-go theorem for M/superstring theory could be removed. Firstly, the accelerated expansion needs a violation of the strong-energy condition, which requires that $R_{00} \geq 0$ under normal circumstances. However $R_{00}$ must be negative to accommodate to the acceleration of a FLRW (homogeneous and isotropic) universe. Secondly, many four dimensional supergravity theories violated the strong energy condition, except eleven dimensional supergravity or any of the ten dimensional supergravity theories that serve as effective field theories for a superstring theory. The third observation is that if the higher-dimensional stress tensor satisfies the strong energy condition then so will the lower dimensional stress tensor and is applicable to the types of compactification mentioned in [3].

In the paper, they showed there exist cosmological compactifications on Einstein spaces of negative curvature that can give an accelerating four-dimensional FLRW cosmology in Einstein frame. Their vacuum solutions could produce accelerating cosmologies if one assumes compact hyperbolic internal space that is time dependent. The space-time, considered is in the form of $R^{3+1} \times H_n$.

## 2.5 A Note on Acceleration from Product Space Compactification

Chen et al. [5] have derived a general formulation of Einstein equations for vacuum solutions for product spaces of flat, spherical and hyperbolic spaces. They also examine a simpler case, which is a product of two compact spaces, $R^{3+1} \times K_1 \times K_2$. However, these cases could not give sufficient inflation, but they showed that in principle, one could get an eternal inflation from the similar models with only coefficients of order one. They also analyzed general product spaces $K_0 \times K_1 \times ...... \times K_n$, where the 0-sector of space does not need to be flat and its number of dimensions does not have to be 3. They then obtained the exact solutions for the product spaces in which $K_1 = K_2 = ...... = K_n$, where $K_i$ can be a spherical, hyperbolic or flat space.

As a summary, they generalized the Townsend and Wohlfarth's work in Ref. [3] and examined the compactifications of product spaces of flat, spherical and hyperbolic spaces. Finally, the acceleration phases are still not sufficient in all the cases they have examined.

## 2.6 Hyperbolic Space Cosmologies

Chen et al. have investigated the models compactified on a product of hyperbolic and flat spaces [8] in order to get a model that give sufficient inflation. Generally, solutions are difficult to obtain for general product space compactifications. Therefore, they try to use a specific ansatz and then found a solution for a new class of vacuum space-times, which is a product of flat and hyperbolic spaces.

In addition to the above, they also considered solutions for the cases of hyperbolic external dimensions, external hyperbolic space with internal flat space, and the case of both external and internal spaces being hyperbolic with the hope of getting a sufficient inflation. In the final case, it gives solutions whose late time behavior is like a Milne space-time, expanding with constant rate. From perturbative expansions of such Milne solutions, they found that eternally accelerating expansion can be obtained for which the number of internal dimensions is greater than or equal to seven.

# CHAPTER 3

# METHODOLOGY AND THEORY

## 3.1 Einstein's Field Equation

In 1915, Albert Einstein developed his general theory of relativity. The important equation which play the main role in this theory is what we call Einstein's field equation or just Einstein's equation:

$$R_{\mu\nu} - \frac{1}{2} g_{\mu\nu} R = \frac{8\pi G}{c^4} T_{\mu\nu},\qquad(3.1\text{-}1)$$

where $R_{\mu\nu}$ = Ricci tensor,

$g_{\mu\nu}$ = Metric tensor,

$T_{\mu\nu}$ = Energy-momentum tensor,

$R$ = Ricci scalar.

Einstein's field equation determines how the metric (left hand side) responds to energy and momentum of matter. This metric dually describes the geometry of space-time and the gravitational field of the matter source. The solutions of the Einstein's field equation are metrics of space-time.

The equation (3.1-1) can also be rewritten as

$$R_{\mu\nu} = \frac{8\pi G}{c^4}\left(T_{\mu\nu} - \frac{1}{2} T g_{\mu\nu}\right),\qquad(3.1\text{-}2)$$

where $T$ = Trace of the energy-momentum tensor, to substantiate the second perspective.

### 3.1.1 The Cosmological Constant

When Einstein developed his equation of general relativity in 1915, the expansion of the universe had not been discovered. So the people at that time believed that the universe was static. Therefore, Einstein modified his equation by adding a term proportional to the metric in order to get a static solution.

Equation (3.1-1) now become:

$$R_{\mu\nu} - \frac{1}{2} g_{\mu\nu} R - \Lambda g_{\mu\nu} = \frac{8\pi G}{c^4} T_{\mu\nu}, \qquad (3.1.1\text{-}1)$$

where $\Lambda$ = cosmological constant,

$G$ = Newton's gravitational constant.

In 1929, Edwin Hubble made an observation and found that the universe is expanding. For this reason, Einstein regretted adding the cosmological constant into his field equations and so he subsequently ignored this term.

### 3.1.2 Vacuum Solutions of the Field Equation

If the energy –momentum tensor, $T_{\mu\nu}$ is zero in the region under consideration, then the solution to

$$R_{\mu\nu} = 0 \qquad (3.1.2\text{-}1)$$

is referred to as the vacuum solution.

If the cosmological constant is nonzero, the vacuum solution is the solution to

$$R_{\mu\nu} - \frac{1}{2} R g_{\mu\nu} = \Lambda g_{\mu\nu} \quad . \tag{3.1.2-2}$$

One example of (3.1.2-1) is Schwarzschild solution:

$$c^2 d\tau^2 = \left(1 - \frac{2MG}{c^2 r}\right) c^2 dt^2 - \left(1 - \frac{2MG}{c^2 r}\right)^{-1} dr^2 - r^2 d\theta^2 - r^2 \sin^2\theta d\phi^2, \tag{3.1.2-3}$$

where $M$ = Mass of the matter.

This solution is very useful in describing the gravitational force outside a spherical body. Another way of representing the equation (3.1.2-1) is through the variation of Einstein-Hilbert action:

$$S = k \int R \sqrt{-g} d^n x \quad , \tag{3.1.2-4}$$

where $k = \dfrac{c^4}{16\pi G}$,

$g = \det(g_{\mu\nu})$,

$R$ = Ricci scalar.

### 3.1.3 Non-vacuum Solutions of the Field Equation

One of the important examples is Robertson-Walker metric:

$$ds^2 = c^2 dt^2 - R^2(t) \left[ \frac{dr^2}{1 - kr^2} + r^2 \left( d\theta^2 + \sin^2\theta d\phi^2 \right) \right], \tag{3.1.3-1}$$

where $R(t)$ = scale factor

$k$ = -1, 0, +1.

We substitute the metric (3.1.3-1) into Einstein's equation (3.1-1) in order to get the behavior of the scale factor. The resultant equations are called Friedmann equations. These Friedmann equations relate the scale factor to the energy-momentum of the universe. The matter and energy we consider here is of the perfect fluid.

Usually, we consider

$$\left(\frac{\dot{R}}{R}\right)^2 = \frac{8\pi G \varepsilon}{3c^2} - \frac{c^2 k}{R^2} \qquad (3.1.3\text{-}2)$$

as the Friedmann equation, where $\varepsilon$ and $p$ is the energy density and pressure respectively, and

$$\frac{\ddot{R}}{R} = -\frac{4\pi G}{3c^2}(\varepsilon + 3p) \qquad (3.1.3\text{-}3)$$

as the *second* Friedmann equation.

If the cosmological constant is nonzero, the equation (3.1.3-2) becomes:

$$\left(\frac{\dot{R}}{R}\right)^2 = \frac{8\pi G \varepsilon}{3c^2} - \frac{c^2 k}{R^2} + \frac{c^2 \Lambda}{3} \quad, \qquad (3.1.3\text{-}4)$$

and equation (3.1.3-3) becomes:

$$\frac{\ddot{R}}{R} = -\frac{4\pi G}{3c^2}(\varepsilon + 3p) + \frac{c^2 \Lambda}{3} \quad. \qquad (3.1.3\text{-}5)$$

### 3.1.4 Some Important Formulas Related to Einstein's Equation

As we have already discussed in the last section, the solution of the Einstein's field equation is a metric. In what follows are some of the important equations related to the metrics in general relativity.

(1) Christoffel symbols:

$$\Gamma^{\mu}_{\nu\lambda} = \frac{1}{2} g^{\mu\sigma} \left( g_{\sigma\nu,\lambda} + g_{\sigma\lambda,\nu} - g_{\nu\lambda,\sigma} \right). \tag{3.1.4-1}$$

The interpretation of $\Gamma^{\mu}_{\nu\lambda}$ is that it is the $\mu$-th component of $\frac{\partial \vec{e}_\nu}{\partial x^\lambda}$, $\left( \frac{\partial \vec{e}_\nu}{\partial x^\lambda} = \Gamma^{\mu}_{\nu\lambda} \vec{e}_\mu \right)$, where $\vec{e}_\nu$ and $\vec{e}_\mu$ is basis vector. It is constructed from the components of metric tensor.

(2) Riemann tensor:

$$R^{\sigma}_{\mu\nu\lambda} = \Gamma^{\sigma}_{\mu\lambda,\nu} - \Gamma^{\sigma}_{\mu\nu,\lambda} + \Gamma^{\sigma}_{\alpha\nu}\Gamma^{\alpha}_{\mu\lambda} - \Gamma^{\sigma}_{\alpha\lambda}\Gamma^{\alpha}_{\mu\nu}. \tag{3.1.4-2}$$

This is a mathematical description of the intrinsic curvature of a manifold. It is constructed completely from the Christoffel symbols. If a coordinate system exists in which the components of the metric are constant, the Riemann tensor will vanish.

(3) Ricci tensor:

$$R_{\mu\nu} = \Gamma^{\lambda}_{\mu\nu,\lambda} - \Gamma^{\lambda}_{\mu\lambda,\nu} + \Gamma^{\lambda}_{\mu\nu}\Gamma^{\sigma}_{\lambda\sigma} - \Gamma^{\sigma}_{\mu\lambda}\Gamma^{\lambda}_{\nu\sigma}. \tag{3.1.4-3}$$

The Ricci tensor is essentially the only contraction of the Riemann tensor, $R_{\mu\lambda} = R^{\nu}_{\mu\nu\lambda}$. The Ricci tensor is symmetric as a consequence of the symmetries of the Riemann tensor, $R_{\mu\lambda} = R_{\lambda\mu}$.

(4) Ricci scalar:

$$R = g^{\mu\nu} R_{\mu\nu}. \tag{3.1.4-4}$$

The Ricci scalar is the trace of the Ricci tensor, $R = R^{\mu}_{\mu} = g^{\mu\nu} R_{\mu\nu}$.

(5) Energy-momentum tensor:

$$T_{\mu\nu} = (\varepsilon + p) u_\mu u_\nu - p g_{\mu\nu}, \tag{3.1.4-5}$$

where $u_\mu$ is the four-velocity of matter.

This energy-momentum tensor is for the model of perfect fluids. A perfect fluid is one that can be completely specified by two quantities, the rest-frame energy density $\rho$, and an isotropic rest-frame pressure $p$. The concept of a perfect fluid is general enough to describe a wide variety of physical forms of matter. Dust is a special case for which $p = 0$.

## 3.2   Accelerating Cosmology with Dark Energy

If we consider our universe on the very large scale, it is spatially flat. The WMAP satellite confirmed this recently [21,22,23] by measuring the cosmic microwave background (CMB). In the case of flat universe, the density must equal to a critical density. This critical density is proportional to the square of the Hubble constant and is about $9 \times 10^{-27}$ kilograms per cubic meter [24]. But the problem is that from observation, the amount of matter, including baryons and dark matter in our universe is just about 30% of the critical density. Therefore, a hypothetical form of energy (or mass), is thought to be smoothly distributed throughout the universe and has strong negative pressure. This is consistent with the discovery by the Supernova Cosmology Project (SCP) [25] and the High-Z Supernova team [26] that the expansion of universe is accelerating. Accelerating universe requires a type of energy that has negative pressure and this pressure could act to speed up the expansion. This form of energy is called dark energy (the term dark energy was coined by Michael Turner) [24]. Dark energy does not emit light. The two candidates for dark energy are cosmological constant and quintessence. The equation of state parameter, $w$ of dark energy is $w < -\frac{1}{3}$. Dark energy is roughly having density of $10^{-29}$ grams per cubic centimeter.

### 3.2.1 Behavior of Dark Energy

Einstein introduced the cosmological constant, $\Lambda$ in 1917 in order to get static solution. Actually, this parameter not only allows us to get a static universe but also decelerating and accelerating universe. The cosmological constant may be a positive value, zero or negative value. Its dimensionality is $(length)^{-2}$. As shown in Ref. [24], Einstein added this parameter in his field equation to get the static solution. Firstly, we consider the case in Newtonian context.

There is an equation, called Poisson's equation:

$$\nabla^2 \Phi = 4\pi G \varepsilon , \qquad (3.2.1\text{-}1)$$

where $\Phi$ = gravitational potential and $\varepsilon$ = density of the universe.

Gravitational acceleration is then the gradient of the potential:

$$\vec{a} = -\vec{\nabla}\Phi . \qquad (3.2.1\text{-}2)$$

In a static universe, $\vec{a} = 0$ at every point of universe. So, we conclude for every point of universe, $\Phi$ = constant.

We can also rearrange the equation (3.2.1-1) to be

$$\varepsilon = \frac{1}{4\pi G} \nabla^2 \Phi \qquad (3.2.1\text{-}3)$$

If $\Phi$ is constant, then $\nabla^2\Phi = 0$. So, we conclude that the universe is static if the universe is empty. We know that this situation is impossible. Therefore, Einstein added a constant to his field equation (3.1-1) and later was called cosmological constant. In Newtonian context, the equation (3.2.1-1) now becomes:

$$\nabla^2\Phi + \Lambda = 4\pi G\varepsilon . \qquad (3.2.1\text{-}4)$$

So, now we conclude that the universe is static if $\Lambda = 4\pi G\varepsilon$.

In general relativity context, we have in the subsection 3.1.3, the equation (3.1.3-5):

$$\frac{\ddot{R}}{R} = -\frac{4\pi G}{3c^2}(\varepsilon + 3p) + \frac{c^2\Lambda}{3} . \qquad (3.2.1\text{-}5)$$

From this equation, we can conclude that the universe is accelerating if:

$$\Lambda > \frac{4\pi G(\varepsilon + 3p)}{c^4} . \qquad (3.2.1\text{-}6)$$

We can also relook at equation (3.1.3-4):

$$\left(\frac{\dot{R}}{R}\right)^2 = \frac{8\pi G\varepsilon}{3c^2} - \frac{c^2 k}{R^2} + \frac{c^2\Lambda}{3} . \qquad (3.2.1\text{-}7)$$

From (3.2.1-7), the cosmological constant is just like another component's energy density. Usually, it is also called vacuum energy density according to quantum mechanics. Using such perspective we can write

$$\Lambda = \frac{8\pi G}{c^4}\varepsilon_\Lambda \qquad (3.2.1\text{-}8)$$

where $\varepsilon_\Lambda$ = vacuum energy density corresponding to cosmological constant.

Before we write it out as an equation of state, $p = w\varepsilon$, we have to consider the following first.

Conservation of energy-momentum tensor gives:

$$\dot{\varepsilon} + 3\frac{\dot{R}}{R}(\varepsilon + p) = 0 \quad . \tag{3.2.1-9}$$

We know from equation (3.2.1-8) that $\varepsilon_\Lambda$ is constant with time. So, equation (3.2.1-9) reduces to:

$$(\varepsilon_\Lambda + p_\Lambda) = 0$$

$$p_\Lambda = -\varepsilon_\Lambda. \tag{3.2.1-10}$$

Now, we can make the conclusion that the cosmological constant, $\Lambda$ is having an equation of state parameter, $w = -1$.

The other candidate is quintessence [24]; it originates from theoretical high-energy physics. It is also a hypothetical form of dark energy and can explain why the universe is accelerating. Quintessence is a dynamic field, which is time evolving, and spatially dependent. The equation of state parameter of quintessence lies in the range of $-1 < w < -\frac{1}{3}$. However, there is still no strong evidence for existence of quintessence.

### 3.2.2 Some Issues About Dark Energy

From quantum field theory, the calculated value for vacuum energy (dark energy) density is about $10^{112} erg/cm^3$ [27]. But now, we have an observation of vacuum energy density is $10^{-8} erg/cm^3$. This is clearly a big discrepancy between the theoretical value and observed value. This problem is called cosmological constant problem. Besides this, there is also a coincidence problem. The problem refers to why the dark energy density is approximately equal to the current matter density. This problem can be seen from the following.

From [27], we have:

$$\frac{\Omega_\Lambda}{\Omega_M} = \frac{\varepsilon_\Lambda}{\varepsilon_M} \propto R(t)^3, \qquad (3.2.2\text{-}1)$$

where $\varepsilon_M$ = matter energy density in the universe,

$\Omega_\Lambda$ = Density parameter for vacuum energy,

$\Omega_M$ = Density parameter for matter.

This is showing that the ratio of these energy density changes rapidly as the universe expands.

## 3.3   Method

In this work, most calculations are based on Einstein's field equation. Einstein's field equation is made up of Ricci tensor, metric tensor, Ricci scalar and energy-momentum tensor. The general idea in this work is that by defining a proper space-time metric in higher dimensions, calculating respective tensors and then substitute into Einstein's equation. A solution, which can describe the accelerating universe, is then sought for. These solutions will usually present a relation between the scale factor and the matter contained in the universe. We also consider the empty universe, whose energy-momentum tensor is zero.

One of the important calculations before solving Einstein's equation is the computation of Christoffel symbol. Fortunately, the metric tensors that are considered here are only having the non-zero components in the diagonal. This will reduce down the number of the non-zero Christoffel symbols (equation 3.1.4-1). They are:

$$\Gamma^{\mu}_{\mu\nu} = \frac{1}{2} g^{\mu\mu} \left( g_{\mu\mu,\nu} \right),$$

$$\Gamma^{\mu}_{\nu\mu} = \frac{1}{2} g^{\mu\mu} \left( g_{\mu\mu,\nu} \right),$$

$$\Gamma^{\mu}_{\nu\nu} = \frac{1}{2} g^{\mu\mu} \left( -g_{\nu\nu,\mu} \right). \tag{3.3-1}$$

After having a list of non-zero Christoffel symbols, we can calculate the Ricci tensor by using equation (3.1.4-3). We also can calculate the Ricci scalar with equation (3.1.4-4) and finally we get the Einstein tensor,

$$G_{\mu\nu} = R_{\mu\nu} - \frac{1}{2} g_{\mu\nu} R. \tag{3.3-2}$$

After this, we can proportionately equate the Einstein tensor with energy-momentum tensor with the coupling constant, $\frac{8\pi G}{c^4}$. The Einstein tensor is equated to zero if the space-time is empty.

One should consider that Einstein's equations are not constrained by the number of dimensions; the number of dimensions is actually arbitrary. Hence, in this work, we well consider the Einstein equation in higher dimensions.

In the section 4.1, the work is on the generalization of the Friedmann equation and will reduce to the usual Friedmann equation if we consider zero extra dimensions. While in the section 4.2, we consider the extra dimension in the context of supergravity that is the low energy limit of superstring theory. Here, we reduce the $4+n$-dimensional action into $4D$ Einstein-Hilbert action plus the action for the scalar fields by introducing the Einstein frame metric. The scalar fields are responsible for inflation.

For the section 4.1, the solutions are just the higher dimensional Friedmann equation and in which we apply the observational data into the solutions and then assume some physical behavior of the extra dimensions in order to check what is the spatial curvature of the internal spaces. Even the way of interpretation is simple, but it gives a similar physical behavior for the extra dimensions in the work of section 4.2. The work of section 4.1 is a slight generalization of those in [17,18,19,20], instead of a single internal space; here we consider the extra dimensions to be products of many spaces. The final solutions are in a general form.

While in the section 4.2, the theory is from the supergravity point of view, which serves as effective field theory for a superstring/M theory. For supergravity theory, the point particle is considered; but in superstring or M theory, this point particle is corresponding to string or membrane. In other word, supergravity theory is just a low energy limit of superstring/M theory. This can be understood when the string or membrane is considered as having zero length or zero area.

The work in section 4.2 is also just slight generalization of [8]. Here, the internal spaces are products of two flat spaces and one hyperbolic space; whereas in [8], the internal spaces are products of one flat space and one hyperbolic space. The metric considered here must be in the form of Einstein frame metric in order to ensure that gravity part of the action is in the Einstein-Hilbert form, which means that the Ricci scalar is not multiplied by an additional term. The existence of this additional term indicates that the effective Newton constant becomes time dependent. This is not preferable for the model of universe that is considered here. Then, the scale factor in term of t is determined from vacuum solution and Einstein frame metric. From the scale factor, there is a time interval in which the universe is undergoing accelerated expansion.

# CHAPTER 4

# CALCULATION AND RESULTS

## 4.1　Field's Equation with Extra Dimensions

Today, many cosmological models explain that the present accelerating expansion universe is caused by a type of hidden energy and is so called "dark energy" which is distributed in the three dimensional spatial universe homogeneously. In this section we will derive Einstein's field equation in higher dimensions with the extra dimensions as products of $n$ spaces. These spaces are spatial part of the Robertson-Walker metric. Each space can be flat, spherical or hyperbolic which we denote it by a numbers $k_i$, which can be 0, +1 or –1 respectively. In this extra dimensional point of view, we will assume that this extra dimensions is actually corresponding to the dark energy. The argument is that according to the theory of relativity, the mass (energy) is related to the geometry of space-time intimately. In our case, the dark energy will behave just like the matter and cause the small regions of space curved throughout the whole universe. This curved space caused by the dark energy is actually extra dimensions. On the other hand, we also can say that the extra dimensions contain a type of energy, which is called dark energy. Therefore, we can argue that not only the energy can cause the space-time to curve, but when there is an extra dimension, it must install a mysterious energy. The more detailed can be found in Ref. [28].

In this section, we will derive the Einstein's equation in higher dimensions, and the extra dimensional part is a product of $n$ spaces.

This space-time is having the form as follows:

$$R \times M_0 \times M_1 \times M_2 \times M_3 \times \ldots \ldots \times M_n ,\qquad(4.1\text{-}1)$$

where the $M_0$ is our ordinary three dimensions space.

We also assume that the matter content in this higher dimensional universe is a perfect fluid with the energy-momentum tensor given by:

$$T^\mu_\nu = diag\left(\varepsilon, -p_0, -p_0, -p_0, \ldots, -p_1, -p_1, \ldots, -p_2, -p_2, \ldots\right). \qquad(4.1\text{-}2)$$

The space-time is described by the following metric:

$$ds^2 = dt^2 - \sum_{i=0}^{n}\left[R_i^2\left(\frac{dr_i^2}{1-k_i r_i^2} + r_i^2 d\Omega_d^{(i)2}\right)\right], \qquad(4.1\text{-}3)$$

where $i = 0,1,2,3,\ldots\ldots,n$;

$$d\Omega_d^{(i)2} = d\theta_1^{(i)2} + \sin^2\theta_1^{(i)} d\theta_2^{(i)2} + \ldots\ldots + \sin^2\theta_1^{(i)}\sin^2\theta_2^{(i)}\ldots\ldots\sin^2\theta_{d-1}^{(i)} d\theta_d^{(i)2};$$

number of dimensions for the $i$-th space, $m_i = d_i + 1$.

The metric components are listed as follows:

$$g_{tt} = 1,\ g_{r_i r_i} = \frac{-R_i^2}{\left(1-k_i r_i^2\right)},\ g_{\theta_1^{(i)}\theta_1^{(i)}} = -R_i^2 r_i^2,\ g_{\theta_2^{(i)}\theta_2^{(i)}} = -R_i^2 r_i^2 \sin^2\theta_1^{(i)},$$

$$g_{\theta_s^{(i)}\theta_s^{(i)}} = -R_i^2 r_i^2 \sin^2\theta_1^{(i)} \sin^2\theta_2^{(i)} \ldots\ldots \sin^2\theta_{s-1}^{(i)},$$

$$g^{tt} = 1,\ g^{r_i r_i} = -\frac{\left(1-k_i r_i^2\right)}{R_i^2},\ g^{\theta_1^{(i)}\theta_1^{(i)}} = -\frac{1}{R_i^2 r_i^2},\ g^{\theta_2^{(i)}\theta_2^{(i)}} = -\frac{1}{R_i^2 r_i^2 \sin^2\theta_1^{(i)}},$$

$$g^{\theta_s^{(i)}\theta_s^{(i)}} = -\frac{1}{R_i^2 r_i^2 \sin^2\theta_1^{(i)} \sin^2\theta_2^{(i)} \ldots \sin^2\theta_{s-1}^{(i)}}. \tag{4.1-4}$$

The nonzero Christoffel symbols can be computed by using the equation (3.1.4-1) and are given as follows:

$$\Gamma^t_{r_i r_i} = \frac{R_i \dot{R}_i}{(1-k_i r_i^2)}, \quad \Gamma^t_{\theta_1^{(i)}\theta_1^{(i)}} = R_i \dot{R}_i r_i^2, \quad \Gamma^t_{\theta_2^{(i)}\theta_2^{(i)}} = R_i \dot{R}_i r_i^2 \sin^2\theta_1^{(i)},$$

$$\Gamma^t_{\theta_s^{(i)}\theta_s^{(i)}} = R_i \dot{R}_i r_i^2 \sin^2\theta_1^{(i)} \sin^2\theta_2^{(i)} \ldots \sin^2\theta_{s-1}^{(i)},$$

$$\Gamma^{r_i}_{t r_i} = \frac{\dot{R}_i}{R_i}, \quad \Gamma^{r_i}_{r_i r_i} = \frac{k_i r_i}{(1-k_i r_i^2)}, \quad \Gamma^{r_i}_{\theta_1^{(i)}\theta_1^{(i)}} = -r_i(1-k_i r_i^2),$$

$$\Gamma^{r_i}_{\theta_2^{(i)}\theta_2^{(i)}} = -r_i(1-k_i r_i^2)\sin^2\theta_1^{(i)},$$

$$\Gamma^{r_i}_{\theta_s^{(i)}\theta_s^{(i)}} = -r_i(1-k_i r_i^2)\sin^2\theta_1^{(i)} \sin^2\theta_2^{(i)} \ldots \sin^2\theta_{s-1}^{(i)}$$

$$\Gamma^{\theta_1^{(i)}}_{t\theta_1^{(i)}} = \frac{\dot{R}_i}{R_i}, \quad \Gamma^{\theta_1^{(i)}}_{r_i\theta_1^{(i)}} = \frac{1}{r_i}, \quad \Gamma^{\theta_{1_i}}_{\theta_2^{(i)}\theta_2^{(i)}} = -\sin\theta_1^{(i)} \cos\theta_1^{(i)},$$

$$\Gamma^{\theta_2^{(i)}}_{t\theta_2^{(i)}} = \frac{\dot{R}_i}{R_i}, \quad \Gamma^{\theta_2^{(i)}}_{r_i\theta_2^{(i)}} = \frac{1}{r_i}, \quad \Gamma^{\theta_2^{(i)}}_{\theta_1^{(i)}\theta_2^{(i)}} = \cot\theta_1^{(i)},$$

$$\Gamma^{\theta_s^{(i)}}_{t\theta_s^{(i)}} = \frac{\dot{R}_i}{R_i}, \quad \Gamma^{\theta_s^{(i)}}_{r_i\theta_s^{(i)}} = \frac{1}{r_i},$$

$$\Gamma^{\theta_p^{(i)}}_{\theta_q^{(i)}\theta_q^{(i)}} = -\frac{\left(\sin^2\theta_1^{(i)} \sin^2\theta_2^{(i)} \ldots \sin^2\theta_{q-1}^{(i)}\right)\sin\theta_p^{(i)}\cos\theta_p^{(i)}}{\left(\sin^2\theta_1^{(i)} \sin^2\theta_2^{(i)} \ldots \sin^2\theta_p^{(i)}\right)}, \quad (q>p)$$

$$\Gamma^{\theta_q^{(i)}}_{\theta_p^{(i)}\theta_q^{(i)}} = \cot\theta_p^{(i)}, \quad (q>p), \tag{4.1-5}$$

where $s = 1, 2, 3, \ldots$; $p = 1, 2, 3, \ldots$; $q = 1, 2, 3, \ldots$.

Next, the Christoffel symbols are substituted into (3.1.4-3) to get the nonzero components of the Ricci tensor.

The tt-component of Ricci tensor is

$$R_{tt} = -\Gamma^{\lambda}_{t\lambda,t} - \Gamma^{\sigma}_{t\lambda}\Gamma^{\lambda}_{t\sigma}$$

$$= -\Gamma^{r_0}_{tr_0,t} - \Gamma^{\theta^{(0)}_1}_{t\theta^{(0)}_1,t} - \Gamma^{\theta^{(0)}_2}_{t\theta^{(0)}_2,t} - \cdots\cdots - \Gamma^{r_1}_{tr_1,t} - \Gamma^{\theta^{(1)}_1}_{t\theta^{(1)}_1,t} - \Gamma^{\theta^{(1)}_2}_{t\theta^{(1)}_2,t} - \cdots\cdots$$

$$-\Gamma^{r_2}_{tr_2,t} + \Gamma^{\theta^{(2)}_1}_{t\theta^{(2)}_1,t} + \Gamma^{\theta^{(2)}_2}_{t\theta^{(2)}_2,t} - \cdots\cdots - \Gamma^{r_0}_{tr_0}\Gamma^{r_0}_{tr_0} - \Gamma^{\theta^{(0)}_1}_{t\theta^{(0)}_1}\Gamma^{\theta^{(0)}_1}_{t\theta^{(0)}_1} - \Gamma^{\theta^{(0)}_2}_{t\theta^{(0)}_2}\Gamma^{\theta^{(0)}_2}_{t\theta^{(0)}_2} - \cdots\cdots$$

$$-\Gamma^{r_1}_{tr_1}\Gamma^{r_1}_{tr_1} - \Gamma^{\theta^{(1)}_1}_{t\theta^{(1)}_1}\Gamma^{\theta^{(1)}_1}_{t\theta^{(1)}_1} - \Gamma^{\theta^{(1)}_2}_{t\theta^{(1)}_2}\Gamma^{\theta^{(1)}_2}_{t\theta^{(1)}_2} - \cdots\cdots - \Gamma^{r_2}_{tr_2}\Gamma^{r_2}_{tr_2} - \Gamma^{\theta^{(2)}_1}_{t\theta^{(2)}_1}\Gamma^{\theta^{(2)}_1}_{t\theta^{(2)}_1} - \Gamma^{\theta^{(2)}_2}_{t\theta^{(2)}_2}\Gamma^{\theta^{(2)}_2}_{t\theta^{(2)}_2} - \cdots\cdots$$

$$= -\sum_{i=0}^{n} m_i \left( \frac{\ddot{R}_i R_i - \dot{R}_i^2}{R_i^2} \right) - \sum_{i=0}^{n} m_i \frac{\dot{R}_i^2}{R_i^2}$$

$$= -\sum_{i=0}^{n} m_i \frac{\ddot{R}_i}{R_i}. \qquad (4.1\text{-}6)$$

The $r_i r_i$-component of Ricci tensor is

$$R_{r_i r_i} = \Gamma^{\lambda}_{r_i r_i,\lambda} - \Gamma^{\lambda}_{r_i \lambda,r_i} + \Gamma^{\lambda}_{r_i r_i}\Gamma^{\sigma}_{\lambda\sigma} - \Gamma^{\sigma}_{r_i \lambda}\Gamma^{\lambda}_{r_i \sigma}$$

$$= \Gamma^{t}_{r_i r_i,t} - \Gamma^{\theta^{(i)}_1}_{r_i \theta^{(i)}_1,r_i} - \Gamma^{\theta^{(i)}_2}_{r_i \theta^{(i)}_2,r_i} - \cdots\cdots + \Gamma^{t}_{r_i r_i}\Gamma^{r_0}_{tr_0} + \Gamma^{t}_{r_i r_i}\Gamma^{\theta^{(0)}_1}_{t\theta^{(0)}_1} + \Gamma^{t}_{r_i r_i}\Gamma^{\theta^{(0)}_2}_{t\theta^{(0)}_2} + \cdots\cdots$$

$$+ \Gamma^{t}_{r_i r_i}\Gamma^{r_1}_{tr_1} + \Gamma^{t}_{r_i r_i}\Gamma^{\theta^{(1)}_1}_{t\theta^{(1)}_1} + \Gamma^{t}_{r_i r_i}\Gamma^{\theta^{(1)}_2}_{t\theta^{(1)}_2} + \cdots\cdots + \Gamma^{t}_{r_i r_i}\Gamma^{r_2}_{tr_2} + \Gamma^{t}_{r_i r_i}\Gamma^{\theta^{(2)}_1}_{t\theta^{(2)}_1} + \Gamma^{t}_{r_i r_i}\Gamma^{\theta^{(2)}_2}_{t\theta^{(2)}_2} + \cdots\cdots + \cdots\cdots +$$

$$\Gamma^{r_i}_{r_i r_i}\Gamma^{r_i}_{r_i r_i} + \Gamma^{r_i}_{r_i r_i}\Gamma^{\theta^{(i)}_1}_{r_i \theta^{(i)}_1} + \Gamma^{r_i}_{r_i r_i}\Gamma^{\theta^{(i)}_2}_{r_i \theta^{(i)}_2} + \cdots\cdots - \Gamma^{r_i}_{r_i t}\Gamma^{t}_{r_i r_i} - \Gamma^{t}_{r_i r_i}\Gamma^{r_i}_{r_i t} - \Gamma^{r_i}_{r_i r_i}\Gamma^{r_i}_{r_i r_i} - \Gamma^{\theta^{(i)}_1}_{r_i \theta^{(i)}_1}\Gamma^{\theta^{(i)}_1}_{r_i \theta^{(i)}_1}$$

$$-\Gamma^{\theta^{(i)}_2}_{r_i \theta^{(i)}_2}\Gamma^{\theta^{(i)}_2}_{r_i \theta^{(i)}_2} - \cdots\cdots$$

$$= \frac{\dot{R}_i^2 + R_i \ddot{R}_i}{Q_0^{(i)}} + \frac{(m_i - 1)}{r_i^2} + m_i \left( \frac{\dot{R}_i^2}{Q_0^{(i)}} \right) + \left( \frac{R_i \dot{R}_i}{Q_0^{(i)}} \right) \sum_{i \neq j} m_j \left( \frac{\dot{R}_j}{R_j} \right)$$

$$+\frac{k_i^2 r_i^2}{Q_0^{(i)2}} + (m_i - 1)\left(\frac{k_i}{Q_0^{(i)}}\right) - 2\frac{\dot{R}_i^2}{Q_0^{(i)}} - \frac{k_i^2 r_i^2}{Q_0^{(i)2}} - (m_i - 1)\left(\frac{1}{r_i}\right)^2$$

$$= (m_i - 1)\left(\frac{\dot{R}_i^2}{Q_0^{(i)}}\right) + \frac{R_i \ddot{R}_i}{Q_0^{(i)}} + \left(\frac{R_i \dot{R}_i}{Q_0^{(i)}}\right)\sum_{i \neq j} m_j \left(\frac{\dot{R}_j}{R_j}\right) + (m_i - 1)\left(\frac{k_i}{Q_0^{(i)}}\right). \tag{4.1-7}$$

Note that $Q_0^{(i)} = 1 - k_i r_i^2$, $Q_1^{(i)} = r_i^2$, $Q_2^{(i)} = r_i^2 \sin^2 \theta_1^{(i)}$, $Q_s^{(i)} = r_i^2 \sin^2 \theta_1^{(i)} \ldots \sin^2 \theta_{s-1}^{(i)}$.

We know from many textbooks of general relativity or cosmology, for examples [13,29] that:

$$R_{rr} = \frac{W_0}{Q_0}, \quad R_{\theta_1 \theta_1} = Q_1 W_1, \quad R_{\theta_2 \theta_2} = Q_2 W_2, \text{ for which } W_0 = W_1 = W_2.$$

In our case, we have already calculated $R_{r_i r_i}$, and so we does not need to calculate $R_{\theta_s^{(i)} \theta_s^{(i)}}$. Since we have $W_0^{(i)}$, and $W_0^{(i)} = W_1^{(i)} = W_2^{(i)} = W_s^{(i)}$.

Now, we have enough information to compute the Ricci scalar, $R$:

$$R = g^{\mu\nu} R_{\mu\nu}$$

$$= -\sum_{i=0}^{n} m_i \frac{\ddot{R}_i}{R_i} + \sum_{i=0}^{n} \frac{m_i}{-R_i^2}\left[(m_i - 1)\left(\dot{R}_i^2\right) + R_i \ddot{R}_i + \left(R_i \dot{R}_i\right)\sum_{j \neq i}\left(m_j \frac{\dot{R}_j}{R_j}\right) + (m_i - 1) k_i\right]$$

$$= -\sum_{i=0}^{n} m_i \frac{\ddot{R}_i}{R_i} - \sum_{i=0}^{n}\left[\frac{m_i (m_i - 1)\dot{R}_i^2}{R_i^2} + \frac{m_i \ddot{R}_i}{R_i} + \frac{m_i \dot{R}_i}{R_i}\sum_{j \neq i}\left(m_j \frac{\dot{R}_j}{R_j}\right) + \frac{m_i (m_i - 1) k_i}{R_i^2}\right]$$

$$= -2\sum_{i=0}^{n} m_i \frac{\ddot{R}_i}{R_i} - \sum_{i=0}^{n} \frac{m_i (m_i - 1)\dot{R}_i^2}{R_i^2} - \sum_{i=0}^{n} \frac{m_i (m_i - 1) k_i}{R_i^2} - \sum_{i=0}^{n}\left(\frac{m_i \dot{R}_i}{R_i}\sum_{j \neq i} m_j \frac{\dot{R}_j}{R_j}\right). \tag{4.1-8}$$

Now, we can compute the tt-component of the Einstein's equation,

$$R_{tt} - \frac{1}{2} g_{tt} R = 8\pi \bar{G} T_{tt} \qquad \text{(take } c = 1\text{)}$$

$$-\sum_{i=0}^{n}\left(m_i \frac{\ddot{R}_i}{R_i}\right) - \frac{1}{2}\left[-2\sum_{i=0}^{n} m_i \frac{\ddot{R}_i}{R_i} - \sum_{i=0}^{n} \frac{m_i(m_i-1)\dot{R}_i^2}{R_i^2} - \sum_{i=0}^{n} \frac{m_i(m_i-1)k_i}{R_i^2} - \sum_{i=0}^{n}\left(\frac{m_i \dot{R}_i}{R_i} \sum_{j\neq i} m_j \frac{\dot{R}_j}{R_j}\right)\right]$$

$$= 8\pi \bar{G} \varepsilon$$

$$-\sum_{i=0}^{n} m_i \frac{\ddot{R}_i}{R_i} + \sum_{i=0}^{n} m_i \frac{\ddot{R}_i}{R_i} + \frac{1}{2}\sum_{i=0}^{n} \frac{m_i(m_i-1)\dot{R}_i^2}{R_i^2} + \frac{1}{2}\sum_{i=0}^{n} \frac{m_i(m_i-1)k_i}{R_i^2} + \frac{1}{2}\sum_{i=0}^{n}\left(\frac{m_i \dot{R}_i}{R_i} \sum_{j\neq i} m_j \frac{\dot{R}_j}{R_j}\right)$$

$$= 8\pi \bar{G} \varepsilon$$

$$\boxed{\sum_{i=0}^{n}\left\{\frac{m_i(m_i-1)}{2}\left[\left(\frac{\dot{R}_i}{R_i}\right)^2 + \frac{k_i}{R_i^2}\right]\right\} + \sum_{i=0}^{n}\left(\frac{m_i \dot{R}_i}{2R_i} \sum_{j\neq i} m_j \frac{\dot{R}_j}{R_j}\right) = 8\pi \bar{G} \varepsilon} \qquad (4.1\text{-}9)$$

where $\bar{G}$ = higher dimensional Newton's gravitational constant.

We also can compute the Einstein's equation for $r_i r_i$-component and below we will show that the Einstein's equation for $r_i r_i$-component is actually equivalent to all other $\theta_s^{(i)} \theta_s^{(i)}$-component. The $r_i r_i$-component of Einstein's equation is:

$$R_{r_i r_i} - \frac{1}{2} g_{r_i r_i} R = 8\pi \bar{G} T_{r_i r_i}$$

$$(m_i-1)\left(\frac{\dot{R}_i^2}{Q_0^{(i)}}\right) + \frac{R_i \ddot{R}_i}{Q_0^{(i)}} + \left(\frac{R_i \dot{R}_i}{Q_0^{(i)}}\right)\sum_{j\neq i}\left(m_j \frac{\dot{R}_j}{R_j}\right) + (m_i-1)\left(\frac{k_i}{Q_0^{(i)}}\right)$$

$$-\frac{1}{2}\left(-\frac{R_i^2}{Q_0^{(i)}}\right)\left[-2\sum_{i=0}^{n} m_i \frac{\ddot{R}_i}{R_i} - \sum_{i=0}^{n} \frac{m_i(m_i-1)\dot{R}_i^2}{R_i^2} - \sum_{i=0}^{n} \frac{m_i(m_i-1)k_i}{R_i^2} - \sum_{i=0}^{n}\left(\frac{m_i \dot{R}_i}{R_i} \sum_{j\neq i} m_j \frac{\dot{R}_j}{R_j}\right)\right]$$

$$= \frac{8\pi \bar{G} p_i R_i^2}{Q_0^{(i)}}$$

$$\frac{1}{Q_0^{(i)}}\left[(m_i-1)\dot{R}_i^2 + R_i\ddot{R}_i + R_i\dot{R}_i\sum_{j\neq i}m_j\frac{\dot{R}_j}{R_j} + (m_i-1)k_i\right]$$

$$+\frac{R_i^2}{2Q_0^{(i)}}\left[-2\sum_{i=0}^n m_i\frac{\ddot{R}_i}{R_i} - \sum_{i=0}^n \frac{m_i(m_i-1)\dot{R}_i^2}{R_i^2} - \sum_{i=0}^n \frac{m_i(m_i-1)k_i}{R_i^2} - \sum_{i=0}^n\left(\frac{m_i\dot{R}_i}{R_i}\sum_{j\neq i}m_j\frac{\dot{R}_j}{R_j}\right)\right] = \frac{8\pi \bar{G} p_i R_i^2}{Q_0^{(i)}}$$

We note that the $Q_0^{(i)}$ term at each side cancels each other; therefore it does not depend on the $Q$ term. So, the $r_i r_i$-component of Einstein's equation is now obviously equal to all other $\theta_s^{(i)}\theta_s^{(i)}$-component.

$$-\frac{2(m_i-1)\dot{R}_i^2}{2R_i^2} + \frac{m_i(m_i-1)\dot{R}_i^2}{2R_i^2} + \sum_{j\neq i}\frac{m_j(m_j-1)\dot{R}_j^2}{2R_j^2} - \frac{2(m_i-1)k_i}{2R_i^2} + \frac{m_i(m_i-1)k_i}{2R_i^2}$$

$$+\sum_{j\neq i}\frac{m_j(m_j-1)k_j}{2R_j^2} - \frac{\ddot{R}_i}{R_i} + m_i\frac{\ddot{R}_i}{R_i} + \sum_{j\neq i}m_j\frac{\ddot{R}_j}{R_j} - \frac{2\dot{R}_i}{2R_i}\sum_{j\neq i}m_j\frac{\dot{R}_j}{R_j} + \frac{m_i\dot{R}_i}{2R_i}\sum_{j\neq i}m_j\frac{\dot{R}_j}{R_j}$$

$$+\sum_{j\neq i}\left(\frac{m_j\dot{R}_j}{2R_j}\sum_{k\neq j}m_k\frac{\dot{R}_k}{R_k}\right) = -8\pi\bar{G}p_i$$

$$\frac{(m_i-1)(m_i-2)\dot{R}_i^2}{2R_i^2} + \sum_{j\neq i}\frac{m_j(m_j-1)\dot{R}_j^2}{2R_j^2} + \frac{(m_i-1)(m_i-2)k_i}{2R_i^2} + \sum_{j\neq i}\frac{m_j(m_j-1)k_j}{2R_j^2}$$

$$+\frac{(m_i-1)\ddot{R}_i}{R_i} + \sum_{j\neq i}m_j\frac{\ddot{R}_j}{R_j} + (m_i-2)\frac{\dot{R}_i}{2R_i}\sum_{j\neq i}m_j\frac{\dot{R}_j}{R_j} + \sum_{j\neq i}\left(\frac{m_j\dot{R}_j}{2R_j}\sum_{k\neq j}\frac{m_k\dot{R}_k}{R_k}\right) = -8\pi\bar{G}p_i$$

$$\boxed{\begin{aligned}
&(m_i-1)\left(\frac{\ddot{R}_i}{R_i}\right)+\sum_{j\neq i}\left(m_j\frac{\ddot{R}_j}{R_j}\right)+\frac{(m_i-1)(m_i-2)}{2}\left[\left(\frac{\dot{R}_i}{R_i}\right)^2+\left(\frac{k_i}{R_i^2}\right)\right]\\
&+\sum_{j\neq i}\left\{\frac{m_j(m_j-1)}{2}\left[\left(\frac{\dot{R}_j}{R_j}\right)^2+\frac{k_j}{R_j^2}\right]\right\}+\frac{(m_i-2)}{2}\left(\frac{\dot{R}_i}{R_i}\right)\sum_{j\neq i}m_j\frac{\dot{R}_j}{R_j}+\sum_{j\neq i}\left(\frac{m_j\dot{R}_j}{2R_j}\sum_{k\neq j}m_k\frac{\dot{R}_k}{R_k}\right)\\
&=-8\pi\bar{G}p_i
\end{aligned}}$$

(4.1-10)

where $i\in\{0,1,2,3,......,n\}$; $j\in\{0,1,2,3,......,n\}$; $k\in\{0,1,2,3,......,n\}$.

The equation (4.1-9) and (4.1-10) are Einstein's equation corresponding to the higher dimensions.

### 4.1.1 Investigating the Einstein's Equations in Higher Dimensions

Recent measurements of the cosmic microwave background showed that our universe is spatially flat, and Supernova Project WMAP measurements also indicated that the expansion of the present universe is accelerating, $\ddot{R}_0>0$. We therefore set the curvature of the three dimensional ordinary space to be zero, $k_0=0$ and $\ddot{R}_0>0$. Besides this, we also assume that the scale factors of the extra dimensions have the same physical behavior with the scale factor of the ordinary three dimensions space, which are $\ddot{R}_1,\ddot{R}_2,......>0$. Obviously, this is just an assumption because there is no theoretical or observational argument against having $\ddot{R}_1,\ddot{R}_2,......>0$. Here, we also assume that our universe is matter-dominated $(p_0=0)$, regardless of whether there is a dark energy or not, with the perspective that the dark energy may be an effective implication of the existence of the extra dimensions.

**4.1.1.1** $R \times M_{3,k_0} \times M_{m,k_1}$; $m_0 = 3$, $m_1 = m$.

The metric for this type of space-time is as follows:

$$ds^2 = dt^2 - R_0^2(t)\left(\frac{dr_0^2}{1-k_0 r_0^2} + r_0^2 d\Omega_0^2\right) - R_1^2(t)\left(\frac{dr_1^2}{1-k_1 r_1^2} + r_1^2 d\Omega_1^2\right). \quad (4.1.1.1\text{-}1)$$

The 0-th sector of product spaces is ordinary 3-space and 1-th sector of product spaces is the extra dimensional space. The equations (4.1-9) and (4.1-10) now become (some detailed calculations are given in Appendix):

$$3\left[\left(\frac{\dot{R}_0}{R_0}\right)^2 + \frac{k_0}{R_0^2}\right] + \frac{m(m-1)}{2}\left[\left(\frac{\dot{R}_1}{R_1}\right)^2 + \frac{k_1}{R_1^2}\right] + 3m\frac{\dot{R}_0}{R_0}\frac{\dot{R}_1}{R_1} = 8\pi\bar{G}\varepsilon \; ; \quad (4.1.1.1\text{-}2)$$

$$2\frac{\ddot{R}_0}{R_0} + m\frac{\ddot{R}_1}{R_1} + \left[\left(\frac{\dot{R}_0}{R_0}\right)^2 + \frac{k_0}{R_0^2}\right] + \frac{m(m-1)}{2}\left[\left(\frac{\dot{R}_1}{R_1}\right)^2 + \frac{k_1}{R_1^2}\right] + 2m\frac{\dot{R}_0}{R_0}\frac{\dot{R}_1}{R_1} = -8\pi\bar{G}p_0 \; ; \quad (4.1.1.1\text{-}3)$$

$$3\frac{\ddot{R}_0}{R_0} + (m-1)\frac{\ddot{R}_1}{R_1} + 3\left[\left(\frac{\dot{R}_0}{R_0}\right)^2 + \frac{k_0}{R_0^2}\right] + \frac{(m-1)(m-2)}{2}\left[\left(\frac{\dot{R}_1}{R_1}\right)^2 + \frac{k_1}{R_1^2}\right] + 3(m-1)\frac{\dot{R}_0}{R_0}\frac{\dot{R}_1}{R_1}$$

$$= -8\pi\bar{G}p_1. \quad (4.1.1.1\text{-}4)$$

These equations were exactly considered in [17,18,19,20]. However, we will investigate the behavior of the universe from these equations differently. For a matter-dominated universe, $p_0 = 0$ and the ordinary three dimensions space is spatially flat, $k_0 = 0$, the equation (4.1.1.1-3) becomes:

$$2\frac{\ddot{R}_0}{R_0} + \left[\left(\frac{\dot{R}_0}{R_0}\right)^2\right] = -m\frac{\ddot{R}_1}{R_1} - \frac{m(m-1)}{2}\left[\left(\frac{\dot{R}_1}{R_1}\right)^2 + \frac{k_1}{R_1^2}\right] - 2m\frac{\dot{R}_0}{R_0}\frac{\dot{R}_1}{R_1}. \qquad (4.1.1.1\text{-}5)$$

Note that the left hand side is having the positive value and also $\ddot{R}_1, \dot{R}_1 > 0$, which are already assumed in section 4.1.1, so now we are able to conclude for the type of spatial curvature of extra dimensions. We know that the second term on the right hand side must be positive in order to produce the net positive value on the right hand side, therefore

$$-\frac{m(m-1)}{2}\left[\left(\frac{\dot{R}_1}{R_1}\right)^2 + \frac{k_1}{R_1^2}\right] > 0$$

$$\left(\frac{\dot{R}_1}{R_1}\right)^2 + \frac{k_1}{R_1^2} < 0$$

$$k_1 < 0. \qquad (4.1.1.1\text{-}6)$$

We can therefore conclude that the spatial curvature of extra dimensions is negative. This seems to coincide with the superstring/M-theory point of view proposed by Townsend and Wohlfarth for which the internal space is hyperbolic [3].

**4.1.1.2** $R \times M_{3,k_0} \times M_{m_1,k_1} \times M_{m_2,k_2} \times M_{m_3,k_3}$

The metric for this type of space-time is as follows:

$$ds^2 = dt^2 - R_0^2(t)\left(\frac{dr_0^2}{1-k_0 r_0^2} + r_0^2 d\Omega_0^2\right) - R_1^2(t)\left(\frac{dr_1^2}{1-k_1 r_1^2} + r_1^2 d\Omega_1^2\right)$$

$$- R_2^2(t)\left(\frac{dr_2^2}{1-k_2 r_2^2} + r_2^2 d\Omega_2^2\right) - R_3^2(t)\left(\frac{dr_3^2}{1-k_3 r_3^2} + r_3^2 d\Omega_3^2\right). \quad (4.1.1.2\text{-}1)$$

Once again, we set $p_0 = 0, k_0 = 0; \ddot{R}_0, \ddot{R}_1, \ddot{R}_2, \ddot{R}_3 > 0$. We also set $k_1 = k_2 = 0$ but the $k_3$ remains unknown and will be determined in the similar way as in the last subsection. From equation (4.1-10), the Einstein's equation for 0-th sector of product spaces for the above metric is given as follows:

$$2\left(\frac{\ddot{R}_0}{R_0}\right) + \left(\frac{\dot{R}_0}{R_0}\right)^2$$

$$= -\frac{m_3(m_3-1)}{2}\left[\left(\frac{\dot{R}_3}{R_3}\right)^2 + \frac{k_3}{R_3^2}\right] - \frac{m_1(m_1-1)}{2}\left(\frac{\dot{R}_1}{R_1}\right)^2 - \frac{m_2(m_2-1)}{2}\left(\frac{\dot{R}_2}{R_2}\right)^2 - \frac{1}{2}\left(\frac{\dot{R}_0}{R_0}\right)\sum_{j=1}^{3} m_j \frac{\dot{R}_j}{R_j}$$

$$-\sum_{j=1}^{3}\left(\frac{m_j \dot{R}_j}{2R_j}\sum_{k \neq j} m_k \frac{\dot{R}_k}{R_k}\right) - \sum_{j=1}^{3}\left(m_j \frac{\ddot{R}_j}{R_j}\right).$$

(4.1.1.2-2)

From this equation, we know that the left hand side is having the positive value, so it must also be for the right hand side. So, we conclude that the first term on the right hand side of equation (4.1.1.2-2),

$$-\frac{m_3(m_3-1)}{2}\left[\left(\frac{\dot{R}_3}{R_3}\right)^2+\frac{k_3}{R_3^2}\right]>0$$

$$\left(\frac{\dot{R}_3}{R_3}\right)^2+\frac{k_3}{R_3^2}<0$$

$$k_3<0. \qquad (4.1.1.2\text{-}3)$$

Again, a similar conclusion is obtained with remaining extra dimensional space has to be hyperbolic. This case is similar to the case of superstring inspired cosmology considered below.

## 4.2  Accelerating Cosmologies with Superstring/M Theory

Currently, the expansion of the universe is accelerating and was already confirmed by the past years astronomical observations. Hence, many cosmological models were developed, but it is more worthy of note if this phenomena can be explained by a fundamental theory that incorporates both gravity and the standard model of particle physics, and currently, many efforts are from superstring/M theory. However, this supergravity theory could not generate four-dimensional accelerating solutions due to an assumption on the compactified space. This assumption of the no go theorem is that space of the compactified dimensions are time independent. By assuming that the compactified space to be time dependent, Townsend and Wohlfarth showed that vacuum solutions could generate accelerating cosmologies if the extra space is hyperbolic, $R^{3+1}\times H^m$. However in our work, we will consider the extra dimensions to be more general product spaces.

First of all, we consider the following product space metric ansatz:

$$ds^2 = -e^{2A(t)}dt^2 + \sum_{i=0}^{n} e^{2B_i(t)} ds_i^2 ; \tag{4.2-1}$$

where $ds_i^2 = g_{\mu\nu}^{(i)} dx^\mu dx^\nu$ is the metric for each component of the product space with the following form:

$$ds_i^2 = g_{\mu\nu}^{(i)} dx^\mu dx^\nu = \begin{cases} d\chi^2 + \chi^2 d\Omega_{m_i-1}^2, & k_i = 0, \\ d\chi^2 + \sin^2 \chi d\Omega_{m_i-1}^2, & k_i = 1, \\ d\chi^2 + \sinh^2 \chi d\Omega_{m_i-1}^2, & k_i = -1, \end{cases} \tag{4.2-2}$$

where $\chi$ = radial coordinate.

Now, we investigate these three metrics with a scale factor separately for each product term and obtain the Christoffel symbols.

In the zero curvature case, we consider

$$d\sigma_i^2 = -e^{2A} dt^2 + e^{2B_i} ds_i^2, \tag{4.2-3}$$

where $ds_i^2 = d\chi_i^2 + \chi_i^2 d\Omega_{m_i-1}^{(i)2}$;

$$d\Omega_{m_i-1}^{(i)2} = d\theta_1^{(i)2} + \sin^2 \theta_1^{(i)} d\theta_2^{(i)2} + \ldots\ldots + \sin^2 \theta_1^{(i)} \sin^2 \theta_2^{(i)} \ldots\ldots \sin^2 \theta_{m_i-2}^{(i)} d\theta_{m_i-1}^{(i)2};$$

$m_i$ = number of dimensions of the i-th sector of product space.

Using the equation (3.1.4-1), we list out all the nonzero Christoffel symbols in the following [30]:

$$\Gamma_{tt}^t = \frac{dA}{dt}, \quad \Gamma_{\chi_i \chi_i}^t = e^{2B_i - 2A} \frac{dB_i}{dt}, \quad \Gamma_{\theta_1^{(i)} \theta_1^{(i)}}^t = \chi^2 e^{2B_i - 2A} \frac{dB_i}{dt}, \quad \Gamma_{\theta_2^{(i)} \theta_2^{(i)}}^t = e^{2B_i - 2A} \chi^2 \sin^2 \theta_1^{(i)} \frac{dB_i}{dt},$$

$$\Gamma_{\theta_s^{(i)} \theta_s^{(i)}}^t = \chi^2 e^{2B_i - 2A} \frac{dB_i}{dt} \sin^2 \theta_1^{(i)} \sin^2 \theta_2^{(i)} \ldots\ldots \sin^2 \theta_{s-1}^{(i)},$$

$$\Gamma^{\chi_i}_{\chi_i t} = \frac{dB_i}{dt}, \; \Gamma^{\chi_i}_{\theta_1^{(i)}\theta_1^{(i)}} = -\chi_i, \; \Gamma^{\chi_i}_{\theta_2^{(i)}\theta_2^{(i)}} = -\chi_i \sin^2\theta_1^{(i)}, \; \Gamma^{\chi_i}_{\theta_s^{(i)}\theta_s^{(i)}} = -\chi_i \sin^2\theta_1^{(i)} \sin^2\theta_2^{(i)}......\sin^2\theta_{s-1}^{(i)},$$

$$\Gamma^{\theta_1^{(i)}}_{t\theta_1^{(i)}} = \frac{dB_i}{dt}, \; \Gamma^{\theta_1^{(i)}}_{\chi_i\theta_1^{(i)}} = \frac{1}{\chi_i}, \; \Gamma^{\theta_1^{(i)}}_{\theta_2^{(i)}\theta_2^{(i)}} = -\sin\theta_1^{(i)}\cos\theta_1^{(i)},$$

$$\Gamma^{\theta_2^{(i)}}_{\theta_2^{(i)}t} = \frac{dB_i}{dt}, \; \Gamma^{\theta_2^{(i)}}_{\chi_i\theta_2^{(i)}} = \frac{1}{\chi_i}, \; \Gamma^{\theta_2^{(i)}}_{\theta_1^{(i)}\theta_2^{(i)}} = \cot\theta_1^{(i)},$$

$$\Gamma^{\theta_s^{(i)}}_{\chi_i\theta_s^{(i)}} = \frac{1}{\chi_i}, \; \Gamma^{\theta_s^{(i)}}_{\theta_s^{(i)}t} = \frac{dB_i}{dt},$$

$$\Gamma^{\theta_p^{(i)}}_{\theta_q^{(i)}\theta_q^{(i)}} = -\frac{\left(\sin^2\theta_1^{(i)}\sin^2\theta_2^{(i)}......\sin^2\theta_{q-1}^{(i)}\right)\sin\theta_p^{(i)}\cos\theta_p^{(i)}}{\left(\sin^2\theta_1^{(i)}\sin^2\theta_2^{(i)}......\sin^2\theta_p^{(i)}\right)}, \quad (q > p)$$

$$\Gamma^{\theta_q^{(i)}}_{\theta_p^{(i)}\theta_q^{(i)}} = \cot\theta_p^{(i)}, \quad (q > p). \tag{4.2-4}$$

In the positive curvature case, we consider

$$d\sigma_i^2 = -e^{2A}dt^2 + e^{2B_i}ds_i^2, \tag{4.2-5}$$

where $ds_i^2 = d\chi_i^2 + \sin^2\chi_i d\Omega_{m_i-1}^{(i)2}$;

$$d\Omega_{m_i-1}^{(i)2} = d\theta_1^{(i)2} + \sin^2\theta_1^{(i)}d\theta_2^{(i)2} + ...... + \sin^2\theta_1^{(i)}\sin^2\theta_2^{(i)}......\sin^2\theta_{m_i-2}^{(i)}d\theta_{m_i-1}^{(i)2}.$$

The nonzero Christoffel symbols are as follows:

$$\Gamma^t_{tt} = \frac{dA}{dt}, \; \Gamma^t_{\chi_i\chi_i} = e^{2B_i-2A}\frac{dB_i}{dt}, \; \Gamma^t_{\theta_1^{(i)}\theta_1^{(i)}} = \sin^2\chi_i e^{2B_i-2A}\frac{dB_i}{dt},$$

$$\Gamma^t_{\theta_2^{(i)}\theta_2^{(i)}} = e^{2B_i-2A}\sin^2\chi_i \sin^2\theta_1^{(i)}\frac{dB_i}{dt},$$

$$\Gamma^t_{\theta_s^{(i)}\theta_s^{(i)}} = \sin^2\chi_i e^{2B_i-2A}\frac{dB_i}{dt}\sin^2\theta_1^{(i)}\sin^2\theta_2^{(i)}......\sin^2\theta_{s-1}^{(i)},$$

$$\Gamma^{\chi_i}_{\chi_i t} = \frac{dB_i}{dt}, \; \Gamma^{\chi_i}_{\theta_1^{(i)}\theta_1^{(i)}} = -\sin\chi_i\cos\chi_i, \; \Gamma^{\chi_i}_{\theta_2^{(i)}\theta_2^{(i)}} = -\sin\chi_i\cos\chi_i\sin^2\theta_1^{(i)},$$

$$\Gamma^{\chi_i}_{\theta^{(i)}_s \theta^{(i)}_s} = -\sin \chi_i \cos \chi_i \sin^2 \theta^{(i)}_1 \sin^2 \theta^{(i)}_2 \ldots\ldots \sin^2 \theta^{(i)}_{s-1},$$

$$\Gamma^{\theta^{(i)}_1}_{t\theta^{(i)}_1} = \frac{dB_i}{dt}, \quad \Gamma^{\theta^{(i)}_1}_{\chi_i \theta^{(i)}_1} = \cot \chi_i, \quad \Gamma^{\theta^{(i)}_1}_{\theta^{(i)}_2 \theta^{(i)}_2} = -\sin \theta^{(i)}_1 \cos \theta^{(i)}_1,$$

$$\Gamma^{\theta^{(i)}_2}_{\theta^{(i)}_2 t} = \frac{dB_i}{dt}, \quad \Gamma^{\theta^{(i)}_2}_{\chi_i \theta^{(i)}_2} = \cot \chi_i, \quad \Gamma^{\theta^{(i)}_2}_{\theta^{(i)}_1 \theta^{(i)}_2} = \cot \theta^{(i)}_1,$$

$$\Gamma^{\theta^{(i)}_s}_{\chi_i \theta^{(i)}_s} = \cot \chi_i, \quad \Gamma^{\theta^{(i)}_s}_{\theta^{(i)}_s t} = \frac{dB_i}{dt}, \quad \Gamma^{\theta^{(i)}_q}_{\theta^{(i)}_p \theta^{(i)}_q} = \cot \theta^{(i)}_p, \quad (q > p)$$

$$\Gamma^{\theta^{(i)}_p}_{\theta^{(i)}_q \theta^{(i)}_q} = -\frac{\left(\sin^2 \theta^{(i)}_1 \sin^2 \theta^{(i)}_2 \ldots\ldots \sin^2 \theta^{(i)}_{q-1}\right) \sin \theta^{(i)}_p \cos \theta^{(i)}_p}{\left(\sin^2 \theta^{(i)}_1 \sin^2 \theta^{(i)}_2 \ldots\ldots \sin^2 \theta^{(i)}_p\right)}, \quad (q > p). \tag{4.2-6}$$

In the negative curvature case, we consider

$$d\sigma^2_i = -e^{2A} dt^2 + e^{2B_i} ds^2_i, \tag{4.2-7}$$

where $ds^2_i = d\chi^2_i + \sinh^2 \chi_i d\Omega^{(i)2}_{m_i - 1}$;

$$d\Omega^{(i)2}_{m_i - 1} = d\theta^{(i)2}_1 + \sin^2 \theta^{(i)}_1 d\theta^{(i)2}_2 + \ldots\ldots + \sin^2 \theta^{(i)}_1 \sin^2 \theta^{(i)}_2 \ldots\ldots \sin^2 \theta^{(i)}_{m_i - 2} d\theta^{(i)2}_{m_i - 1}.$$

The nonzero Christoffel symbols are as follows:

$$\Gamma^t_{tt} = \frac{dA}{dt}, \quad \Gamma^t_{\chi_i \chi_i} = e^{2B_i - 2A} \frac{dB_i}{dt}, \quad \Gamma^t_{\theta^{(i)}_1 \theta^{(i)}_1} = \sinh^2 \chi_i e^{2B_i - 2A} \frac{dB_i}{dt},$$

$$\Gamma^t_{\theta^{(i)}_2 \theta^{(i)}_2} = e^{2B_i - 2A} \sinh^2 \chi_i \sin^2 \theta^{(i)}_1 \frac{dB_i}{dt},$$

$$\Gamma^t_{\theta^{(i)}_s \theta^{(i)}_s} = \sinh^2 \chi_i e^{2B_i - 2A} \frac{dB_i}{dt} \sin^2 \theta^{(i)}_1 \sin^2 \theta^{(i)}_2 \ldots\ldots \sin^2 \theta^{(i)}_{s-1},$$

$$\Gamma^{\chi_i}_{\chi_i t} = \frac{dB_i}{dt}, \quad \Gamma^{\chi_i}_{\theta^{(i)}_1 \theta^{(i)}_1} = -\sinh \chi_i \cosh \chi_i, \quad \Gamma^{\chi_i}_{\theta^{(i)}_2 \theta^{(i)}_2} = -\sinh \chi_i \cosh \chi_i \sin^2 \theta^{(i)}_1,$$

$$\Gamma^{\chi_i}_{\theta^{(i)}_s \theta^{(i)}_s} = -\sinh \chi_i \cosh \chi_i \sin^2 \theta^{(i)}_1 \sin^2 \theta^{(i)}_2 \ldots\ldots \sin^2 \theta^{(i)}_{s-1},$$

$$\Gamma^{\theta_1^{(i)}}_{t\theta_1^{(i)}} = \frac{dB_i}{dt}, \quad \Gamma^{\theta_1^{(i)}}_{\chi_i \theta_1^{(i)}} = \coth \chi_i, \quad \Gamma^{\theta_1^{(i)}}_{\theta_2^{(i)} \theta_2^{(i)}} = -\sin \theta_1^{(i)} \cos \theta_1^{(i)},$$

$$\Gamma^{\theta_2^{(i)}}_{\theta_2^{(i)} t} = \frac{dB_i}{dt}, \quad \Gamma^{\theta_2^{(i)}}_{\chi_i \theta_2^{(i)}} = \coth \chi_i, \quad \Gamma^{\theta_2^{(i)}}_{\theta_1^{(i)} \theta_2^{(i)}} = \cot \theta_1^{(i)},$$

$$\Gamma^{\theta_s^{(i)}}_{\chi_i \theta_s^{(i)}} = \coth \chi_i, \quad \Gamma^{\theta_s^{(i)}}_{\theta_s^{(i)} t} = \frac{dB_i}{dt}, \quad \Gamma^{\theta_q^{(i)}}_{\theta_p^{(i)} \theta_q^{(i)}} = \cot \theta_p^{(i)}, \quad (q > p)$$

$$\Gamma^{\theta_p^{(i)}}_{\theta_q^{(i)} \theta_q^{(i)}} = -\frac{\left( \sin^2 \theta_1^{(i)} \sin^2 \theta_2^{(i)} ...... \sin^2 \theta_{q-1}^{(i)} \right) \sin \theta_p^{(i)} \cos \theta_p^{(i)}}{\left( \sin^2 \theta_1^{(i)} \sin^2 \theta_2^{(i)} ...... \sin^2 \theta_p^{(i)} \right)}, \quad (q > p). \tag{4.2-8}$$

Now, we have enough information to compute the Ricci tensor, $R_{tt}, R_{r_i r_i}, R_{\theta_1^{(i)} \theta_1^{(i)}}, ......$

$$R_{tt} = \Gamma^\lambda_{tt,\lambda} - \Gamma^\lambda_{t\lambda,t} + \Gamma^\lambda_{tt} \Gamma^\sigma_{\lambda \sigma} - \Gamma^\sigma_{t\lambda} \Gamma^\lambda_{t\sigma}$$

$$= \Gamma^t_{tt,t} - \Gamma^t_{tt,t} - \Gamma^{\chi_0}_{t\chi_0,t} - \Gamma^{\theta_1^{(0)}}_{t\theta_1^{(0)},t} - \Gamma^{\theta_2^{(0)}}_{t\theta_2^{(0)},t} - ...... - \Gamma^{\chi_1}_{t\chi_1,t} - \Gamma^{\theta_1^{(1)}}_{t\theta_1^{(1)},t} - \Gamma^{\theta_2^{(1)}}_{t\theta_2^{(1)},t} - ......$$

$$- \Gamma^{\chi_2}_{t\chi_2,t} - \Gamma^{\theta_1^{(2)}}_{t\theta_1^{(2)},t} - \Gamma^{\theta_2^{(2)}}_{t\theta_2^{(2)},t} - ...... + \Gamma^t_{tt} \Gamma^t_{tt} + \Gamma^t_{tt} \Gamma^{\chi_0}_{t\chi_0} + \Gamma^t_{tt} \Gamma^{\theta_1^{(0)}}_{t\theta_1^{(0)}} + \Gamma^t_{tt} \Gamma^{\theta_2^{(0)}}_{t\theta_2^{(0)}} + ......$$

$$+ \Gamma^t_{tt} \Gamma^{\chi_1}_{t\chi_1} + \Gamma^t_{tt} \Gamma^{\theta_1^{(1)}}_{t\theta_1^{(1)}} + \Gamma^t_{tt} \Gamma^{\theta_2^{(1)}}_{t\theta_2^{(1)}} + ...... + ...... - \Gamma^t_{tt} \Gamma^t_{tt} - \Gamma^{\chi_0}_{t\chi_0} \Gamma^{\chi_0}_{t\chi_0} - \Gamma^{\theta_1^{(0)}}_{t\theta_1^{(0)}} \Gamma^{\theta_1^{(0)}}_{t\theta_1^{(0)}}$$

$$- \Gamma^{\theta_2^{(0)}}_{t\theta_2^{(0)}} \Gamma^{\theta_2^{(0)}}_{t\theta_2^{(0)}} - ...... - \Gamma^{\chi_1}_{t\chi_1} \Gamma^{\chi_1}_{t\chi_1} - \Gamma^{\theta_1^{(1)}}_{t\theta_1^{(1)}} \Gamma^{\theta_1^{(1)}}_{t\theta_1^{(1)}} - \Gamma^{\theta_2^{(1)}}_{t\theta_2^{(1)}} \Gamma^{\theta_2^{(1)}}_{t\theta_2^{(1)}} - ...... - \Gamma^{\chi_2}_{t\chi_2} \Gamma^{\chi_2}_{t\chi_2}$$

$$- \Gamma^{\theta_1^{(2)}}_{t\theta_1^{(2)}} \Gamma^{\theta_1^{(2)}}_{t\theta_1^{(2)}} - \Gamma^{\theta_2^{(2)}}_{t\theta_2^{(2)}} \Gamma^{\theta_2^{(2)}}_{t\theta_2^{(2)}} - ......$$

$$= \ddot{A} - \ddot{A} - \sum_{i=0}^{n} \left( m_i \ddot{B}_i \right) + \dot{A}^2 + \dot{A} \sum_{i=0}^{n} \left( m_i \dot{B}_i \right) - \dot{A}^2 - \sum_{i=0}^{n} \left( m_i \dot{B}_i^2 \right)$$

$$= -\sum_{i=0}^{n} m_i \left( \ddot{B}_i + \dot{B}_i^2 - \dot{A} \dot{B}_i \right) \tag{4.2-9}$$

$$R_{\chi_i \chi_i} = \Gamma^\lambda_{\chi_i \chi_i, \lambda} - \Gamma^\lambda_{\chi_i \lambda, \chi_i} + \Gamma^\lambda_{\chi_i \chi_i} \Gamma^\sigma_{\lambda \sigma} - \Gamma^\sigma_{\chi_i \lambda} \Gamma^\lambda_{\chi_i \sigma}$$

$$= \Gamma^{t}_{\chi_i\chi_i,t} - \Gamma^{\theta_1^{(i)}}_{\chi_i\theta_1^{(i)},\chi_i} - \Gamma^{\theta_2^{(i)}}_{\chi_i\theta_2^{(i)},\chi_i} - \ldots + \Gamma^{t}_{\chi_i\chi_i}\Gamma^{t}_{tt} + \Gamma^{t}_{\chi_i\chi_i}\Gamma^{\chi_0}_{t\chi_0} + \Gamma^{t}_{\chi_i\chi_i}\Gamma^{\theta_1^{(0)}}_{t\theta_1^{(0)}} + \Gamma^{t}_{\chi_i\chi_i}\Gamma^{\theta_2^{(0)}}_{t\theta_2^{(0)}} + \ldots$$

$$+ \Gamma^{t}_{\chi_i\chi_i}\Gamma^{\chi_1}_{t\chi_1} + \Gamma^{t}_{\chi_i\chi_i}\Gamma^{\theta_1^{(1)}}_{t\theta_1^{(1)}} + \Gamma^{t}_{\chi_i\chi_i}\Gamma^{\theta_2^{(1)}}_{t\theta_2^{(1)}} + \ldots + \Gamma^{t}_{\chi_i\chi_i}\Gamma^{\chi_2}_{t\chi_2} + \Gamma^{t}_{\chi_i\chi_i}\Gamma^{\theta_1^{(2)}}_{t\theta_1^{(2)}}$$

$$+ \Gamma^{t}_{\chi_i\chi_i}\Gamma^{\theta_2^{(2)}}_{t\theta_2^{(2)}} + \ldots + \ldots - \Gamma^{\chi_i}_{\chi_i t}\Gamma^{t}_{\chi_i\chi_i} - \Gamma^{t}_{\chi_i\chi_i}\Gamma^{\chi_i}_{\chi_i t} - \Gamma^{\theta_1^{(i)}}_{\chi_i\theta_1^{(i)}}\Gamma^{\theta_1^{(i)}}_{\chi_i\theta_1^{(i)}} - \Gamma^{\theta_2^{(i)}}_{\chi_i\theta_2^{(i)}}\Gamma^{\theta_2^{(i)}}_{\chi_i\theta_2^{(i)}} - \ldots$$

$$= \Gamma^{t}_{\chi_i\chi_i,t} - \Gamma^{\theta_1^{(i)}}_{\chi_i\theta_1^{(i)},\chi_i} - \Gamma^{\theta_2^{(i)}}_{\chi_i\theta_2^{(i)},\chi_i} - \ldots + \Gamma^{t}_{\chi_i\chi_i}\Gamma^{t}_{tt} + \Gamma^{t}_{\chi_i\chi_i}\sum_{i=0}^{n}\left(m_i\dot{B}_i\right) - 2\Gamma^{t}_{\chi_i\chi_i}\Gamma^{\chi_i}_{\chi_i t}$$

$$- (m_i - 1)\left(\Gamma^{\theta_s^{(i)}}_{\chi_i\theta_s^{(i)}}\right)^2$$

$$= e^{2B_i - 2A}\left(\ddot{B}_i - \dot{A}\dot{B}_i + \dot{B}_i\sum_{j=0}^{n}m_j\dot{B}_j\right) - (m_i - 1)\left[\Gamma^{\theta_s^{(i)}}_{\chi_i\theta_s^{(i)},\chi_i} + \left(\Gamma^{\theta_s^{(i)}}_{\chi_i\theta_s^{(i)}}\right)^2\right] \qquad (4.2\text{-}10)$$

$$R_{\theta_1^{(i)}\theta_1^{(i)}} = \Gamma^{\lambda}_{\theta_1^{(i)}\theta_1^{(i)},\lambda} - \Gamma^{\lambda}_{\theta_1^{(i)}\lambda,\theta_1^{(i)}} + \Gamma^{\lambda}_{\theta_1^{(i)}\theta_1^{(i)}}\Gamma^{\sigma}_{\lambda\sigma} - \Gamma^{\sigma}_{\theta_1^{(i)}\lambda}\Gamma^{\lambda}_{\theta_1^{(i)}\sigma}$$

$$= \Gamma^{t}_{\theta_1^{(i)}\theta_1^{(i)},t} + \Gamma^{\chi_i}_{\theta_1^{(i)}\theta_1^{(i)},\chi_i} - \Gamma^{\theta_2^{(i)}}_{\theta_1^{(i)}\theta_2^{(i)},\theta_1^{(i)}} - \Gamma^{\theta_3^{(i)}}_{\theta_1^{(i)}\theta_3^{(i)},\theta_1^{(i)}} - \ldots + \Gamma^{t}_{\theta_1^{(i)}\theta_1^{(i)}}\Gamma^{t}_{tt} + \Gamma^{t}_{\theta_1^{(i)}\theta_1^{(i)}}\Gamma^{\chi_0}_{t\chi_0} + \Gamma^{t}_{\theta_1^{(i)}\theta_1^{(i)}}\Gamma^{\theta_1^{(0)}}_{t\theta_1^{(0)}}$$

$$+ \Gamma^{t}_{\theta_1^{(i)}\theta_1^{(i)}}\Gamma^{\theta_2^{(0)}}_{t\theta_2^{(0)}} + \ldots + \Gamma^{t}_{\theta_1^{(i)}\theta_1^{(i)}}\Gamma^{\chi_1}_{t\chi_1} + \Gamma^{t}_{\theta_1^{(i)}\theta_1^{(i)}}\Gamma^{\theta_1^{(1)}}_{t\theta_1^{(1)}} + \Gamma^{t}_{\theta_1^{(i)}\theta_1^{(i)}}\Gamma^{\theta_2^{(1)}}_{t\theta_2^{(1)}} + \ldots$$

$$+ \Gamma^{t}_{\theta_1^{(i)}\theta_1^{(i)}}\Gamma^{\chi_2}_{t\chi_2} + \Gamma^{t}_{\theta_1^{(i)}\theta_1^{(i)}}\Gamma^{\theta_1^{(2)}}_{t\theta_1^{(2)}} + \Gamma^{t}_{\theta_1^{(i)}\theta_1^{(i)}}\Gamma^{\theta_2^{(2)}}_{t\theta_2^{(2)}} + \ldots + \ldots + \Gamma^{\chi_i}_{\theta_1^{(i)}\theta_1^{(i)}}\Gamma^{\theta_1^{(i)}}_{\chi_i\theta_1^{(i)}}$$

$$+ \Gamma^{\chi_i}_{\theta_1^{(i)}\theta_1^{(i)}}\Gamma^{\theta_2^{(i)}}_{\chi_i\theta_2^{(i)}} + \ldots - \Gamma^{t}_{\theta_1^{(i)}\theta_1^{(i)}}\Gamma^{\theta_1^{(i)}}_{\theta_1^{(i)}t} - \Gamma^{\chi_i}_{\theta_1^{(i)}\theta_1^{(i)}}\Gamma^{\theta_1^{(i)}}_{\theta_1^{(i)}\chi_i} - \Gamma^{\theta_1^{(i)}}_{\theta_1^{(i)}t}\Gamma^{t}_{\theta_1^{(i)}\theta_1^{(i)}} - \Gamma^{\theta_1^{(i)}}_{\theta_1^{(i)}\chi_i}\Gamma^{\chi_i}_{\theta_1^{(i)}\theta_1^{(i)}}$$

$$- \Gamma^{\theta_2^{(i)}}_{\theta_1^{(i)}\theta_2^{(i)}}\Gamma^{\theta_2^{(i)}}_{\theta_1^{(i)}\theta_2^{(i)}} - \Gamma^{\theta_3^{(i)}}_{\theta_1^{(i)}\theta_3^{(i)}}\Gamma^{\theta_3^{(i)}}_{\theta_1^{(i)}\theta_3^{(i)}} - \ldots$$

$$= Q_1^{(i)}e^{2B_i - 2A}\left(2\dot{B}_i^2 - 2\dot{A}\dot{B}_i + \ddot{B}_i\right) + \Gamma^{\chi_i}_{\theta_1^{(i)}\theta_1^{(i)},\chi_i} + \frac{(m_i - 2)}{\sin^2\theta_1^{(i)}} + Q_1^{(i)}e^{2B_i - 2A}\dot{B}_i\dot{A}$$

$$+ Q_1^{(i)}e^{2B_i - 2A}\dot{B}_i\sum_{j=0}^{n}\left(m_j\dot{B}_j\right) + \Gamma^{\chi_i}_{\theta_1^{(i)}\theta_1^{(i)}}(m_i - 1)\Gamma^{\theta_s^{(i)}}_{\chi_i\theta_s^{(i)}} - 2\Gamma^{t}_{\theta_1^{(i)}\theta_1^{(i)}}\Gamma^{\theta_1^{(i)}}_{\theta_1^{(i)}t} - 2\Gamma^{\chi_i}_{\theta_1^{(i)}\theta_1^{(i)}}\Gamma^{\theta_1^{(i)}}_{\theta_1^{(i)}\chi_i}$$

$$- (m_i - 2)\cot^2\theta_1^{(i)}$$

$$= -Q_1^{(i)} e^{2B_i - 2A} \dot{A}\dot{B}_i + Q_1^{(i)} e^{2B_i - 2A} \ddot{B} + Q_1^{(i)} e^{2B_i - 2A} \dot{B}_i \sum_{j=0}^{n} m_j \dot{B}_j + (m_i - 2)$$

$$+ \Gamma^{\chi_i}_{\theta_1^{(i)}\theta_1^{(i)}, \chi_i} + (m_i - 3) \Gamma^{\chi_i}_{\theta_1^{(i)}\theta_1^{(i)}} \Gamma^{\theta_1^{(i)}}_{\chi_i \theta_1^{(i)}} \tag{4.2-11}$$

$$R_{\theta_2^{(i)}\theta_2^{(i)}} = \Gamma^{\lambda}_{\theta_2^{(i)}\theta_2^{(i)}, \lambda} - \Gamma^{\lambda}_{\theta_2^{(i)}\lambda, \theta_2^{(i)}} + \Gamma^{\lambda}_{\theta_2^{(i)}\theta_2^{(i)}} \Gamma^{\sigma}_{\lambda\sigma} - \Gamma^{\sigma}_{\theta_2^{(i)}\lambda} \Gamma^{\lambda}_{\theta_2^{(i)}\sigma}$$

$$= \Gamma^{t}_{\theta_2^{(i)}\theta_2^{(i)}, t} + \Gamma^{\chi_i}_{\theta_2^{(i)}\theta_2^{(i)}, \chi_i} + \Gamma^{\theta_1^{(i)}}_{\theta_2^{(i)}\theta_2^{(i)}, \theta_1^{(i)}} - \Gamma^{\theta_3^{(i)}}_{\theta_2^{(i)}\theta_3^{(i)}, \theta_2^{(i)}} - \Gamma^{\theta_4^{(i)}}_{\theta_2^{(i)}\theta_4^{(i)}, \theta_2^{(i)}} - \ldots\ldots + \Gamma^{t}_{\theta_2^{(i)}\theta_2^{(i)}} \Gamma^{t}_{tt} + \Gamma^{t}_{\theta_2^{(i)}\theta_2^{(i)}} \Gamma^{\chi_0}_{t\chi_0}$$

$$+ \Gamma^{t}_{\theta_2^{(i)}\theta_2^{(i)}} \Gamma^{\theta_1^{(0)}}_{t\theta_1^{(0)}} + \Gamma^{t}_{\theta_2^{(i)}\theta_2^{(i)}} \Gamma^{\theta_2^{(0)}}_{t\theta_2^{(0)}} + \ldots\ldots + \Gamma^{t}_{\theta_2^{(i)}\theta_2^{(i)}} \Gamma^{\chi_1}_{t\chi_1} + \Gamma^{t}_{\theta_2^{(i)}\theta_2^{(i)}} \Gamma^{\theta_1^{(1)}}_{t\theta_1^{(1)}} + \Gamma^{t}_{\theta_2^{(i)}\theta_2^{(i)}} \Gamma^{\theta_2^{(1)}}_{t\theta_2^{(1)}} + \ldots\ldots + \ldots\ldots$$

$$+ \Gamma^{\chi_i}_{\theta_2^{(i)}\theta_2^{(i)}} \Gamma^{\theta_1^{(i)}}_{\chi_i \theta_1^{(i)}} + \Gamma^{\chi_i}_{\theta_2^{(i)}\theta_2^{(i)}} \Gamma^{\theta_2^{(i)}}_{\chi_i \theta_2^{(i)}} + \ldots\ldots + \Gamma^{\theta_1^{(i)}}_{\theta_2^{(i)}\theta_2^{(i)}} \Gamma^{\theta_2^{(i)}}_{\theta_1^{(i)}\theta_2^{(i)}} + \Gamma^{\theta_1^{(i)}}_{\theta_2^{(i)}\theta_2^{(i)}} \Gamma^{\theta_3^{(i)}}_{\theta_1^{(i)}\theta_3^{(i)}} + \ldots\ldots$$

$$- \Gamma^{t}_{\theta_2^{(i)}\theta_2^{(i)}} \Gamma^{\theta_2^{(i)}}_{\theta_2^{(i)} t} - \Gamma^{\chi_i}_{\theta_2^{(i)}\theta_2^{(i)}} \Gamma^{\theta_2^{(i)}}_{\theta_2^{(i)} \chi_i} - \Gamma^{\theta_1^{(i)}}_{\theta_2^{(i)}\theta_2^{(i)}} \Gamma^{\theta_2^{(i)}}_{\theta_2^{(i)}\theta_1^{(i)}} - \Gamma^{\theta_2^{(i)}}_{\theta_2^{(i)} t} \Gamma^{t}_{\theta_2^{(i)}\theta_2^{(i)}} - \Gamma^{\theta_2^{(i)}}_{\theta_2^{(i)} \chi_i} \Gamma^{\chi_i}_{\theta_2^{(i)}\theta_2^{(i)}}$$

$$- \Gamma^{\theta_2^{(i)}}_{\theta_2^{(i)}\theta_1^{(i)}} \Gamma^{\theta_1^{(i)}}_{\theta_2^{(i)}\theta_2^{(i)}} - \Gamma^{\theta_3^{(i)}}_{\theta_2^{(i)}\theta_3^{(i)}} \Gamma^{\theta_3^{(i)}}_{\theta_2^{(i)}\theta_3^{(i)}} - \Gamma^{\theta_4^{(i)}}_{\theta_2^{(i)}\theta_4^{(i)}} \Gamma^{\theta_4^{(i)}}_{\theta_2^{(i)}\theta_4^{(i)}} - \ldots\ldots$$

$$= 2Q_2^{(i)} e^{2B_i - 2A} \dot{B}^2 - 2Q_2^{(i)} e^{2B_i - 2A} \dot{A}\dot{B}_i + Q_2^{(i)} e^{2B_i - 2A} \ddot{B}_i + \Gamma^{\chi_i}_{\theta_2^{(i)}\theta_2^{(i)}, \chi_i} + \sin^2 \theta_1^{(i)} - \cos^2 \theta_1^{(i)}$$

$$+ \frac{(m_i - 3)}{\sin^2 \theta_2^{(i)}} + Q_2^{(i)} e^{2B_i - 2A} \dot{A}\dot{B}_i + Q_2^{(i)} e^{2B_i - 2A} \dot{B}_i \sum_{j=0}^{n} m_j \dot{B}_j + (m_i - 1) \Gamma^{\chi_i}_{\theta_2^{(i)}\theta_2^{(i)}} \Gamma^{\theta_s^{(i)}}_{\chi_i \theta_s^{(i)}}$$

$$+ (m_i - 2)\left(-\sin \theta_1^{(i)} \cos \theta_1^{(i)}\right) \cot \theta_1^{(i)} - Q_2^{(i)} e^{2B_i - 2A} \dot{B}^2$$

$$- \Gamma^{\chi_i}_{\theta_2^{(i)}\theta_2^{(i)}} \Gamma^{\theta_2^{(i)}}_{\theta_2^{(i)} \chi_i} - \left(-\sin \theta_1^{(i)} \cos \theta_1^{(i)}\right) \cot \theta_1^{(i)} - Q_2^{(i)} e^{2B_i - 2A} \dot{B}^2 - \Gamma^{\theta_2^{(i)}}_{\theta_2^{(i)} \chi_i} \Gamma^{\chi_i}_{\theta_2^{(i)}\theta_2^{(i)}}$$

$$- \left(-\sin \theta_1^{(i)} \cos \theta_1^{(i)}\right) \cot \theta_1^{(i)} - (m_i - 3) \cot^2 \theta_2^{(i)}$$

$$= -Q_2^{(i)} e^{2B_i - 2A} \dot{A}\dot{B}_i + Q_2^{(i)} e^{2B_i - 2A} \ddot{B}_i + Q_2^{(i)} e^{2B_i - 2A} \dot{B}_i \sum_{j=0}^{n} m_j \dot{B}_j + 1 + \Gamma^{\chi_i}_{\theta_2^{(i)}\theta_2^{(i)}, \chi_i}$$

$$+ \frac{(m_i - 3)}{\sin^2 \theta_2^{(i)}} + (m_i - 1) \Gamma^{\chi_i}_{\theta_2^{(i)}\theta_2^{(i)}} \Gamma^{\theta_s^{(i)}}_{\chi_i \theta_s^{(i)}} - (m_i - 2) \cos^2 \theta_1^{(i)} - 2\Gamma^{\chi_i}_{\theta_2^{(i)}\theta_2^{(i)}} \Gamma^{\theta_2^{(i)}}_{\theta_2^{(i)} \chi_i}$$

$$- (m_i - 3) \cot^2 \theta_2^{(i)}, \text{ etc.} \tag{4.2-12}$$

After we consider the zero curvature case, positive curvature case and negative curvature case for each of these equations (4.2-10), (4.2-11) and (4.2-12), we get an equation in the general form:

$$R^{(i)}_{\mu\nu} = \left\{ e^{2B_i - 2A}\left[ \ddot{B}_i + \dot{B}_i\left(-\dot{A} + \sum_{j=0}^{n} m_j \dot{B}_j \right) \right] + k_i(m_i - 1) \right\} g^{(i)}_{\mu\nu} \qquad (4.2\text{-}13)$$

Note that

$$g^{(i)}_{\chi\chi} = 1 \quad (k_i = 0, +1, -1),$$

$$g^{(i)}_{\theta_1 \theta_1} = \chi^2 \quad (k_i = 0),$$
$$\qquad = \sin^2 \chi \quad (k_i = +1),$$
$$\qquad = \sinh^2 \chi \quad (k_i = -1),$$

$$g^{(i)}_{\theta_2 \theta_2} = \chi^2 \sin^2 \theta_1^{(i)} \quad (k_i = 0),$$
$$\qquad = \sin^2 \chi \, \sin^2 \theta_1^{(i)} \quad (k_i = +1),$$
$$\qquad = \sinh^2 \chi \sin^2 \theta_1^{(i)} \quad (k_i = -1),$$

$$g^{(i)}_{\theta_s \theta_s} = \chi^2 \sin^2 \theta_1^{(i)} \ldots \sin^2 \theta_{s-1}^{(i)} \quad (k_i = 0),$$
$$\qquad = \sin^2 \chi \, \sin^2 \theta_1^{(i)} \ldots \sin^2 \theta_{s-1}^{(i)} \quad (k_i = +1), \qquad (4.2\text{-}14)$$
$$\qquad = \sinh^2 \chi \sin^2 \theta_1^{(i)} \ldots \sin^2 \theta_{s-1}^{(i)} \quad (k_i = -1).$$

The equation (4.2-13) can be simplified by using the condition:

$$-A + \sum_{j=0}^{n} m_j B_j = 0. \qquad (4.2\text{-}15)$$

This is effectively just a time reparametrization because all the functions depend only on time.

The vacuum Einstein's equations then reduce to:

$$R_{tt} = 0$$

$$-\sum_{i=0}^{n} m_i \ddot{B}_i + \sum_{i=0}^{n} m_i (m_i - 1) \dot{B}_i^2 + \sum_{j \neq i}^{n} m_i m_j \dot{B}_i \dot{B}_j = 0 \qquad (4.2\text{-}16)$$

$$R_{\mu\nu}^{(i)} = 0$$

$$\ddot{B}_i + \in_i (m_i - 1) e^{2(m_i - 1)B_i + 2\sum_{j \neq i} m_j B_j} = 0 \qquad (4.2\text{-}17)$$

### 4.2.1 Corresponding Action

The metric of the space-time considered here also can be written as follows:

$$ds_{4+m}^2 = a^2 ds_4^2 + b_1^2 ds_{m_1}^2 + b_2^2 ds_{m_2}^2 + \ldots\ldots \qquad (4.2.1\text{-}1)$$

The corresponding action is:

$$S = \int d^{4+m} x \sqrt{-g}\, R$$

$$= \int d^m x \sqrt{g_m} \int d^4 x \sqrt{-g_4}\, a^4 \times b^{m_1} \times \ldots\ldots \left[ \frac{R_4}{a^2} + \frac{R_{m_1}}{b^2} + \ldots\ldots \right] \qquad (4.2.1\text{-}2)$$

This action reduces to the sum of the 4d Einstein-Hilbert action plus an action for the scalar field, $\phi_i = \ln b_i$ if $a = \left( \dfrac{1}{b^{m_b} \times \ldots\ldots} \right)^{\frac{1}{2}}$. If $a = \left( \dfrac{1}{b^{m_b} \times \ldots\ldots} \right)^{\frac{1}{2}}$, then the metric (4.2.1-1) is called $4+m$-dimensional Einstein frame metric. In other words, when the Einstein frame metric is substituted into the action (4.2.1-2), the action will reduce down to 4 dimensions. If there is only one internal space like in Ref. [3], then the second term in (4.2.1-2) yields a scalar field potential for $\phi$,

$$V = -R_m e^{-(m+2)\phi} \tag{4.2.1-3}$$

Thus in order to get a positive potential (accelerated expansion), the internal space must be hyperbolic. This is why we have to compactify the internal space with negative curvature. For more details, please see the references [5,31,32].

By referring to the definition of Einstein frame metric above, the metric (4.2-1) in Einstein frame metric is as follows:

$$ds_D^2 = \sum_{i=1}^{n} e^{2B_i(t)} ds_{m_i,\in_i}^2 + e^{-\frac{2}{d-1}\sum_{i=1}^{n} m_i B_i(t)} ds_{E,d+1}^2 \tag{4.2.1-4}$$

and

$$ds_{E,d+1}^2 = e^{\frac{2}{d-1}\sum_{i=1}^{n} m_i B_i} \left( -e^{2\sum_{j=0}^{n} m_j B_j} dt^2 + e^{2B_0} ds_d^2 \right) \tag{4.2.1-5}$$

### 4.2.2 Product Space $R^{3+1} \times R^{m_1} \times R^{m_2} \times H^{m_3}$

This product space is slight generalization of Ref. [8] and is from Ref. [33], so we define similar to Ref. [8],

$$B_0 = \lambda_0 t, \quad B_1 = a - \frac{3\lambda_0 t}{m-1}, \quad B_2 = b - \frac{3\lambda_0 t}{m-1}, \quad B_3 = c - \frac{3\lambda_0 t}{m-1}, \tag{4.2.2-1}$$

where $m = m_1 + m_2 + m_3$, and a, b, c are functions of t. The reason why these functions are defined in such a similar way can be seen from appendix B. Now, we have to solve a, b and c and some steps' detailed calculations will be shown in appendix C.

From equation (4.2-17), we have the following

$$B_0 = \frac{1}{2m_0} \ln\left(\frac{-\ddot{B}_3}{\in_3 (m_3 - 1)}\right) - \frac{(m_3 - 1) B_3 + m_1 B_1 + m_2 B_2}{m_0}; \tag{4.2.2-2}$$

$$\frac{\ddot{B}_1 e^{2B_1}}{k_1 (m_1 - 1)} = \frac{\ddot{B}_2 e^{2B_2}}{k_2 (m_2 - 1)};$$

$$\frac{\ddot{B}_1 e^{2B_1}}{k_1 (m_1 - 1)} = \frac{\ddot{B}_3 e^{2B_3}}{k_3 (m_3 - 1)}; \tag{4.2.2-3}$$

$$\frac{\ddot{B}_2 e^{2B_2}}{k_2 (m_2 - 1)} = \frac{\ddot{B}_3 e^{2B_3}}{k_3 (m_3 - 1)}.$$

From the above equations, (4.2.2-3), (4.2.2-2) and (4.2-16), we have

$$\ddot{a} = 0;$$
$$\ddot{b} = 0; \tag{4.2.2-4}$$

$$\ddot{c} = (m_3 - 1) e^{2c(m_3 - 1) + 2m_1 a + 2m_2 b}; \tag{4.2.2-5}$$

$$-m_3 \ddot{c} + m_1 (m_1 - 1) \dot{a}^2 + m_2 (m_2 - 1) \dot{b}^2 + m_3 (m_3 - 1) \dot{c}^2 + 2m_1 m_2 \dot{a}\dot{b} + 2m_1 m_3 \dot{a}\dot{c}$$
$$+ 2m_2 m_3 \dot{b}\dot{c} = \frac{3\lambda_0^2 (m+2)}{(m-1)}. \tag{4.2.2-6}$$

From equation (4.2.2-4), (4.2.2-5) and (4.2.2-6), we have the following solutions [30,34]:

$$a = \alpha_0 t \; ;$$

$$b = \beta_0 t \; ;$$

$$c = -\frac{m_1 \alpha_0 t}{m_3 - 1} - \frac{m_2 \beta_0 t}{m_3 - 1} + \frac{1}{m_3 - 1} \ln\left(\frac{\sigma}{\sinh[t(m_3 - 1)\sigma]}\right) ; \qquad (4.2.2\text{-}7)$$

$$\sigma^2 = \frac{3\lambda_0^2 (m+2)}{m_3 (m-1)(m_3 - 1)} + \frac{m_1 \alpha_0^2 (m_1 + m_3 - 1)}{m_3 (m_3 - 1)^2} + \frac{m_2 \beta_0^2 (m_2 + m_3 - 1)}{m_3 (m_3 - 1)^2} + \frac{2 m_1 m_2 \alpha_0 \beta_0}{m_3 (m_3 - 1)^2} .$$

Since we already have a, b and c in term of t, we can now evaluate the scale factor. The scale factor can be determined from equation (4.2.1-5). After a short calculation, we can show that

$$ds_{E,d+1}^2 = -S^{2d}(t) dt^2 + S^2(t) ds_{R,d}^2 , \qquad (4.2.2\text{-}8)$$

where $d$ is the number of dimension of our physical universe and is usually taken to be three and $S$ is the scale factor.

If we define $d\tau = S^d dt$, then (4.2.2-8) will become [4]

$$ds_{E,d+1}^2 = -d\tau^2 + S^2(\tau) ds_{R,d}^2 , \qquad (4.2.2\text{-}9)$$

which is the same with the FLRW form.

So, we know that universe is accelerating if

$$\frac{d^2 S}{d\tau^2} > 0 \qquad (4.2.2\text{-}10)$$

Now, the key point is on how we relate the $\alpha_0$ and $\beta_0$ with $\lambda_0$. Here we just assume that there is a value, which is just like coupling constant, relating the $\alpha_0$ and $\beta_0$ with $\lambda_0$. These are:

$$\alpha_0 = c_1 \lambda_0 \quad ; \tag{4.2.2-11}$$

$$\beta_0 = c_2 \lambda_0 \quad . \tag{4.2.2-12}$$

We choose $c_1 = \frac{1}{2}$ and $c_2 = 2$ analogously to the Ref. [8], and are interested in the specific case of $m_1 = 1$, $m_2 = 3$, $m_3 = 3$ with seven dimensional internal space.

Therefore, 
$$\sigma = \frac{\lambda_0 \sqrt{101}}{4} \quad . \tag{4.2.2-13}$$

From equation (4.2.1-5), we find that

$$S = e^{\frac{5\lambda_0 t}{2} + \frac{3c}{2}} \quad . \tag{4.2.2-14}$$

Inequality (4.2.2-10) then leads to the following inequality:

$$\frac{3\sigma^2}{2\sinh^2 2t\sigma} > \left( \frac{19\lambda_0}{8} + \frac{3}{2}\sigma \coth 2t\sigma \right)^2 \quad . \tag{4.2.2-15}$$

The above inequality is satisfied in the interval

$$t_1 = \frac{\ln(0.1197)}{4\sigma} < t < t_2 = \frac{\ln(0.4301)}{4\sigma} \quad . \tag{4.2.2-16}$$

We conclude that by choosing the values $c_1$ and $c_2$ properly and also assuming $\lambda_0$ being a positive value, we can obtain the expansion rate of the internal space, which is ever increasing. If we choose the value of $c_1$ and $c_2$ to be the same, then the expansion rate of the first two internal spaces (flat space) are also the same. This reduces to the case of section 2.4 of Ref. [8]. Therefore, in our case, the first two internal spaces are expanding at different rates even though both of them are flat spaces. This work however also concludes that there is not much difference with the work in [8] in terms of physical features despite the extra product structure and the different scale factors involved.

# CHAPTER 5

# DISCUSSION AND CONCLUSION

The work in section 4.1 is actually studying the Einstein's field equation in the world of higher dimensions. We investigate the implications of accelerating universe on these extra dimensions. By making some assumptions to the extra spaces' scale factors, it is shown that the spatial curvature of extra space, $M_m$ for the space-time $R \times M_3 \times M_m$ is negative. In more generalized space-time of $R \times M_3 \times M_{m_1} \times M_{m_2} \times M_{m_3}$, the spatial curvature of the third component of extra spaces, $M_{m_3}$ is negative if the first and second component of extra spaces are having zero spatial curvature. The assumptions are that the second derivatives of extra dimensional scale factors, $\ddot{R}_1, \ddot{R}_2, \ldots\ldots > 0$. We also assume that the universe is matter dominated, which implies that our universe contains things mostly in matter form. Therefore the pressure can be neglected, $p_0 = 0$. These assumptions lead to the spatial curvature of extra spaces are negative. This conclusion is also coincident with the work of Townsend and Wohlfarth [3] entitled "Accelerating cosmologies from compactification" from the superstring/M theory point of view for which the extra space also must be hyperbolic.

Actually, the work here studies higher dimensional FRW universe and generalizing the extra dimensions part from one space to the product of many spaces, the results are in (4.1-9) and (4.1-10). We do not fix the spatial curvature of the internal space when we start to plug the space-time metric (4.1.1.1-1) and (4.1.1.2-1) into Einstein's field equation. The equations (4.1-9) and (4.1-10) can be reduced to the equations, which were investigated by many authors in their journal papers [17,18,19,20]. From equation (4.1.1.1-5) and (4.1.1.2-2), we conclude that the extra spaces are actually being considered as a source of energy. This energy is able to cause the universe's expansion to accelerate.

In this work, we consider only two types of space-time, which are (4.1.1.1-1) and (4.1.1.2-1). Therefore, there are still many types of space-time yet to be considered. Besides this, it may gives the different results if we change the assumptions or conditions, such as $\ddot{R}$ is not a positive value or we set $k_0 = +1$. We also can investigate the properties of this higher dimensional FRW model, (4.1-9) and (4.1-10) by using the methods that are used in the journal papers [17,18,19,20].

In the section 4.2, we investigate the accelerating cosmologies from string/M theory point of view. We made the extension from the journal paper [8] entitled "Hyperbolic space cosmologies". In that paper, the authors investigate the space-time of $R^{3+1} \times R^{m_1} \times H_{m_2}$. Here, we extend their investigation on the space-time of $R^{3+1} \times R^{m_1} \times R^{m_2} \times H_{m_3}$. The scale factor for each space in the internal space is different with each other. This means that the two flat spaces in the internal space expand with different rate, even though they both are spaces of same curvature and same number of dimension. We also found that this space-time gives accelerating phase in a certain interval of time. For the case studied here, we found that the space-time accelerates in the interval of time, $t_1 = \dfrac{\ln(0.1197)}{4\sigma} < t < t_2 = \dfrac{\ln(0.4301)}{4\sigma}$. If the scale factors for these flat spaces in the internal space are same, then it may yields just like a single flat space in the internal space. While for space-time of $R^{3+1} \times R_m \times R_m \times ...... \times R_m$, $R^{3+1} \times H_m \times H_m \times ...... \times H_m$ and $R^{3+1} \times S_m \times S_m \times ...... \times S_m$, which are investigated in [5], they found that the scale factor for each space in the internal space is the same with each other. Changing the scale factor for the space in internal space of $R^{3+1} \times R^{m_1} \times R^{m_2} \times H_{m_3}$ is able to change the interval of time in which the accelerating phase occurs. This can be done by choosing the different values of parameters in (4.2.2-11) and (4.2.2-12), which are also parts of the equations of scale factors for the space-time of $R^{3+1} \times R^{m_1} \times R^{m_2} \times H_{m_3}$ after dimensional reduction. Even though the interval of time is changed, but this does not seem to change the overall features of the expansion phase.

In short, the works in section 4.1 and 4.2 have similar properties. They both need at least one hyperbolic extra space from the extra dimensional product spaces. Besides this, the extra spaces in both cases are treated as a source of energy, which causes the universe's expansion to accelerate. In section 4.1, the source of energy is treated as dark geometry (or dark energy)[28]. While in section 4.2, the role of the inflation is played by the scalar fields coming from the higher dimensional metric after dimensional reduction. The more detailed can be found in the appendix of Ref. [5].

# APPENDICES

# Appendix A

# Einstein's Equations in Higher Dimensional Universe

(1) Verification of equation (4.1-10) reduces to equation (4.1.1.1-3):

From equation (4.1-10), for $i = 0$,

$$(m_0 - 1)\left(\frac{\ddot{R}_0}{R_0}\right) + \sum_{j \neq 0}\left(m_j \frac{\ddot{R}_j}{R_j}\right) + \frac{(m_0 - 1)(m_0 - 2)}{2}\left[\left(\frac{\dot{R}_0}{R_0}\right)^2 + \left(\frac{k_0}{R_0^2}\right)\right]$$
$$+ \sum_{j \neq 0}\left\{\frac{m_j(m_j - 1)}{2}\left[\left(\frac{\dot{R}_j}{R_j}\right)^2 + \frac{k_j}{R_j^2}\right]\right\} + \frac{(m_0 - 2)}{2}\left(\frac{\dot{R}_0}{R_0}\right)\sum_{j \neq 0} m_j \frac{\dot{R}_j}{R_j} + \sum_{j \neq 0}\left(\frac{m_j \dot{R}_j}{2R_j}\sum_{k \neq j} m_k \frac{\dot{R}_k}{R_k}\right)$$
$$= -8\pi G p_0$$

From $m_0 = 3$ and $m_1 = m$, we have

$$2\left(\frac{\ddot{R}_0}{R_0}\right) + m\frac{\ddot{R}_1}{R_1} + \left[\left(\frac{\dot{R}_0}{R_0}\right)^2 + \left(\frac{k_0}{R_0^2}\right)\right] + \frac{m(m-1)}{2}\left[\left(\frac{\dot{R}_1}{R_1}\right)^2 + \frac{k_1}{R_1^2}\right] + \frac{1}{2}\left(\frac{\dot{R}_0}{R_0}\right)\left(m\frac{\dot{R}_1}{R_1}\right) + \frac{m\dot{R}_1}{2R_1}\left(3\frac{\dot{R}_0}{R_0}\right)$$
$$= -8\pi G p_0$$

$$2\left(\frac{\ddot{R}_0}{R_0}\right) + m\frac{\ddot{R}_1}{R_1} + \left[\left(\frac{\dot{R}_0}{R_0}\right)^2 + \left(\frac{k_0}{R_0^2}\right)\right] + \frac{m(m-1)}{2}\left[\left(\frac{\dot{R}_1}{R_1}\right)^2 + \frac{k_1}{R_1^2}\right] + 2m\left(\frac{\dot{R}_0}{R_0}\right)\left(\frac{\dot{R}_1}{R_1}\right)$$
$$= -8\pi G p_0$$

(AA-1)

The equation (AA-1) is the same with equation (4.1.1.1-3).

(2) Verification of equation (4.1-10) reduces to equation (4.1.1.1-4):

From equation (4.1-10), $i=1$,

$$(m_1-1)\left(\frac{\ddot{R}_1}{R_1}\right)+\sum_{j\neq 1}\left(m_j\frac{\ddot{R}_j}{R_j}\right)+\frac{(m_1-1)(m_1-2)}{2}\left[\left(\frac{\dot{R}_1}{R_1}\right)^2+\left(\frac{k_1}{R_1^2}\right)\right]$$

$$+\sum_{j\neq 1}\left\{\frac{m_j(m_j-1)}{2}\left[\left(\frac{\dot{R}_j}{R_j}\right)^2+\frac{k_j}{R_j^2}\right]\right\}+\frac{(m_1-2)}{2}\left(\frac{\dot{R}_1}{R_1}\right)\sum_{j\neq 1}m_j\frac{\dot{R}_j}{R_j}+\sum_{j\neq 1}\left(\frac{m_j\dot{R}_j}{2R_j}\sum_{k\neq j}m_k\frac{\dot{R}_k}{R_k}\right)$$

$$=-8\pi G p_1$$

From $m_0=3$ and $m_1=m$, we have

$$(m-1)\left(\frac{\ddot{R}_1}{R_1}\right)+3\frac{\ddot{R}_0}{R_0}+\frac{(m-1)(m-2)}{2}\left[\left(\frac{\dot{R}_1}{R_1}\right)^2+\left(\frac{k_1}{R_1^2}\right)\right]+3\left[\left(\frac{\dot{R}_0}{R_0}\right)^2+\frac{k_0}{R_0^2}\right]$$

$$+\frac{(m-2)}{2}\left(\frac{\dot{R}_1}{R_1}\right)\left(3\frac{\dot{R}_0}{R_0}\right)+\frac{3\dot{R}_0}{2R_0}m\frac{\dot{R}_1}{R_1}=-8\pi G p_1$$

$$3\left(\frac{\ddot{R}_0}{R_0}\right)+(m-1)\frac{\ddot{R}_1}{R_1}+3\left[\left(\frac{\dot{R}_0}{R_0}\right)^2+\left(\frac{k_0}{R_0^2}\right)\right]+\frac{(m-1)(m-2)}{2}\left[\left(\frac{\dot{R}_1}{R_1}\right)^2+\frac{k_1}{R_1^2}\right]$$

$$+3(m-1)\left(\frac{\dot{R}_0}{R_0}\right)\left(\frac{\dot{R}_1}{R_1}\right)=-8\pi G p_1$$

(AA-2)

The equation (AA-2) is the same with equation (4.1.1.1-4).

(3) Derivation of equation (4.1.1.2-2):

From equation (4.1-10), for $i = 0$,

$$(m_0 - 1)\left(\frac{\ddot{R}_0}{R_0}\right) + \sum_{j=1}^{3}\left(m_j \frac{\ddot{R}_j}{R_j}\right) + \frac{(m_0 - 1)(m_0 - 2)}{2}\left[\left(\frac{\dot{R}_0}{R_0}\right)^2 + \left(\frac{k_0}{R_0^2}\right)\right]$$

$$+ \sum_{j=1}^{3}\left\{\frac{m_j(m_j - 1)}{2}\left[\left(\frac{\dot{R}_j}{R_j}\right)^2 + \frac{k_j}{R_j^2}\right]\right\} + \frac{(m_0 - 2)}{2}\left(\frac{\dot{R}_0}{R_0}\right)\sum_{j=1}^{3} m_j \frac{\dot{R}_j}{R_j} + \sum_{j=1}^{3}\left(\frac{m_j \dot{R}_j}{2R_j}\sum_{k \neq j} m_k \frac{\dot{R}_k}{R_k}\right)$$

$$= -8\pi \bar{G} p_0$$

From $m_0 = 3, k_0 = 0, k_1 = 0, k_2 = 0, p_0 = 0,$ we have

$$2\left(\frac{\ddot{R}_0}{R_0}\right) + \sum_{j=1}^{3}\left(m_j \frac{\ddot{R}_j}{R_j}\right) + \left(\frac{\dot{R}_0}{R_0}\right)^2 + \frac{m_1(m_1 - 1)}{2}\left(\frac{\dot{R}_1}{R_1}\right)^2 + \frac{m_2(m_2 - 1)}{2}\left(\frac{\dot{R}_2}{R_2}\right)^2$$

$$+ \frac{m_3(m_3 - 1)}{2}\left[\left(\frac{\dot{R}_3}{R_3}\right)^2 + \frac{k_3}{R_3^2}\right] + \frac{1}{2}\left(\frac{\dot{R}_0}{R_0}\right)\sum_{j=1}^{3} m_j \frac{\dot{R}_j}{R_j} + \sum_{j=1}^{3}\left(\frac{m_j \dot{R}_j}{2R_j}\sum_{k \neq j} m_k \frac{\dot{R}_k}{R_k}\right)$$

$$= 0$$

$$2\left(\frac{\ddot{R}_0}{R_0}\right) + \left(\frac{\dot{R}_0}{R_0}\right)^2$$

$$= -\frac{m_1(m_1 - 1)}{2}\left(\frac{\dot{R}_1}{R_1}\right)^2 - \frac{m_2(m_2 - 1)}{2}\left(\frac{\dot{R}_2}{R_2}\right)^2 - \frac{m_3(m_3 - 1)}{2}\left[\left(\frac{\dot{R}_3}{R_3}\right)^2 + \frac{k_3}{R_3^2}\right] - \frac{1}{2}\left(\frac{\dot{R}_0}{R_0}\right)\sum_{j=1}^{3} m_j \frac{\dot{R}_j}{R_j}$$

$$- \sum_{j=1}^{3}\left(\frac{m_j \dot{R}_j}{2R_j}\sum_{k \neq j} m_k \frac{\dot{R}_k}{R_k}\right) - \sum_{j=1}^{3}\left(m_j \frac{\ddot{R}_j}{R_j}\right)$$

(AA-3)

The equation (AA-3) is the same with equation (4.1.1.2-2).

## Appendix B

## Product Space of Same Curvature

It is very difficult to generally solve the coupled differential equations in equations (4.2-16) and (4.2-17). However, we can find exact solutions for some simple cases. Now, I consider the space-time of the form:

$$R^4 \times M_{m_1,k} \times M_{m_2,k} \times M_{m_3,k} \times \ldots \times M_{m_n,k}.  \tag{AB-1}$$

This case was already done in [23], but here I will show the detailed calculations.

From equation (4.2-17), $B_0$ is easily solved and is as follow:

$$B_0 = \alpha_0 t + \beta_0.  \tag{AB-2}$$

We may take $\beta_0 = 0$ by shifting time t. Then, we may define

$$B_i = \frac{-d\alpha_0 t}{m-1} + \sqrt{\frac{m_i - 1}{m-1}} f_i \;,  \tag{AB-3}$$

where $m = \sum_{i=1}^{n} m_i$ ,

$i = 1, 2, 3, \ldots, n$.

Now, we have to find a relation between $B_i$ and $B_k$ $(i \neq k)$. From equation (4.2-17),

$$\ddot{B}_i = -k_i(m_i - 1) e^{2(m_i - 1)B_i + 2\sum_{j \neq i} m_j B_j}$$

$$\ddot{B}_i e^{2B_i} = -k_i(m_i - 1) e^{2\sum_{i=0}^{n} m_i B_i}$$

$$\frac{\ddot{B}_i e^{2B_i}}{k_i(m_i - 1)} = -e^{2A}$$

$$\therefore \quad \frac{\ddot{B}_i e^{2B_i}}{k_i(m_i-1)} = \frac{\ddot{B}_k e^{2B_k}}{k_k(m_k-1)} \tag{AB-4}$$

We know from section 5 of [9], $\ddot{B}_i = \ddot{B}_k$, because the terms of functions of t only involve $m$, but not $m_i$. Therefore, equation (AB-4) becomes

$$e^{2B_k} = \left(\frac{m_k-1}{m_i-1}\right)\frac{k_k e^{2B_i}}{k_i} \tag{AB-5}$$

Then, we can power both sides (AB-5) by $m_k$, it becomes

$$e^{2m_k B_k} = \left(\frac{m_k-1}{m_i-1}\right)^{m_k}\frac{k_k e^{2m_k B_i}}{k_i} \tag{AB-6}$$

In our case, $k_k = k_i$. Hence (AB-6) reduce to

$$e^{2m_k B_k} = \left(\frac{m_k-1}{m_i-1}\right)^{m_k} e^{2m_k B_i} \tag{AB-7}$$

Finally, we get the relation between $B_i$ and $B_k$. This relation is important and will be clear later. From equation (4.2-17),

$$\ddot{B}_i = -k(m_i-1)e^{2(m_i-1)B_i+2\sum_{j\neq i}m_j B_j}$$

$$= -\frac{k(m_i-1)e^{2m_1 B_1} \times e^{2m_2 B_2} \times e^{2m_3 B_3} \times \ldots \times e^{2d\alpha_0 t}}{e^{2B_i}}$$

$$= -k(m_i-1)\prod_{k=1}^{n}\left[\left(\frac{m_k-1}{m_i-1}\right)^{m_k} e^{2m_k B_i}\right] \times e^{-2B_i} \times e^{2d\alpha_0 t}$$

$$= -k(m_i-1)\prod_{k=1}^{n}\left(\frac{m_k-1}{m_i-1}\right)^{m_k} \times e^{2B_i(m_1+m_2+\ldots+m_n)-2B_i} \times e^{2d\alpha_0 t}$$

$$= -k(m_i-1)\prod_{k=1}^{n}\left(\frac{m_k-1}{m_i-1}\right)^{m_k} \times e^{2B_i(m-1)} \times e^{2d\alpha_0 t}$$

$$= -k(m_i - 1) \prod_{k=1}^{n} \left( \frac{m_k - 1}{m_i - 1} \right)^{m_k} \times e^{2(m-1)\left( \frac{-d\alpha_0 t}{m-1} + \sqrt{\frac{m_i-1}{m-1}} f_i \right)} \times e^{2d\alpha_0 t}$$

$$= -k(m_i - 1) \prod_{k=1}^{n} \left( \frac{m_k - 1}{m_i - 1} \right)^{m_k} \times e^{2\sqrt{(m-1)(m_i-1)} f_i} \tag{AB-8}$$

Now, we have to substitute this into the equation (4.2-16). We get

$$\alpha_0^2 d(1-d) + m \left\{ -k(m_i - 1) \prod_{k=1}^{n} \left[ \left( \frac{m_k - 1}{m_i - 1} \right)^{m_k} \right] \times e^{2\sqrt{(m-1)(m_i-1)} f_i} + (1-m) \left( \frac{-d\alpha_0}{m-1} + \sqrt{\frac{m_i-1}{m-1}} \dot{f}_i \right)^2 \right.$$

$$\left. -2d\alpha_0 \left( \frac{-d\alpha_0}{m-1} + \sqrt{\frac{m_i-1}{m-1}} \dot{f}_i \right) \right\} = 0$$

$$\alpha_0^2 d(1-d) - km(m_i - 1) \prod_{k=1}^{n} \left[ \left( \frac{m_k - 1}{m_i - 1} \right)^{m_k} \right] \times e^{2\sqrt{(m-1)(m_i-1)} f_i}$$

$$-m(m-1) \left( \frac{d^2 \alpha_0^2}{(m-1)^2} + \frac{m_i - 1}{m-1} \dot{f}_i^2 - \frac{2d\alpha_0}{m-1} \sqrt{\frac{m_i-1}{m-1}} \dot{f}_i \right) + \frac{2d^2 \alpha_0^2 m}{m-1} - 2dm\alpha_0 \sqrt{\frac{m_i-1}{m-1}} \dot{f}_i = 0$$

$$-km(m_i - 1) \prod_{k=1}^{n} \left[ \left( \frac{m_k - 1}{m_i - 1} \right)^{m_k} \right] \times e^{2\sqrt{(m-1)(m_i-1)} f_i} - m(m_i - 1) \dot{f}_i^2 = -\alpha_0^2 d(1-d) - \frac{d^2 \alpha_0^2 m}{(m-1)}$$

$$k \prod_{k=1}^{n} \left[ \left( \frac{m_k - 1}{m_i - 1} \right)^{m_k} \right] \times e^{2\sqrt{(m-1)(m_i-1)} f_i} + \dot{f}_i^2 = \frac{\alpha_0^2 d(m + d - 1)}{m(m_i - 1)(m - 1)}$$

$$\frac{df_i}{\sqrt{A_i^2 - kC_i e^{2\lambda_i f_i}}} = dt \tag{AB-9}$$

where $A_i^2 = \dfrac{\alpha_0^2 d(m+d-1)}{m(m_i-1)(m-1)}; C_i = \prod_{k=1}^{n}\left(\dfrac{m_k-1}{m_i-1}\right)^{m_k}; \lambda_i = \sqrt{(m-1)(m_i-1)}$.

Now, we have to solve the $f_i$ function. There are 3 cases for 3 different values of $k$, which are $-1, 0, +1$. These 3 cases are considered as follows:

(i) $k = -1$

Equation (AB-9) becomes

$$\frac{df_i}{\sqrt{A_i^2 + C_i e^{2\lambda_i f_i}}} = dt, \tag{AB-10}$$

and then

$$\int \frac{df_i}{\sqrt{A_i^2 + C_i e^{2\lambda_i f_i}}} = \int dt$$

$$\frac{1}{2\lambda_i A_i} \ln\left(\frac{\sqrt{A_i^2 + Ce^{2\lambda_i f_i}} - A_i}{\sqrt{A_i^2 + Ce^{2\lambda_i f_i}} + A_i}\right) = t - t_1 \qquad (t_1 \text{ is integration constant})$$

$$e^{2\lambda_i A_i(t-t_1)} = \frac{\sqrt{A_i^2 + Ce^{2\lambda_i f_i}} - A_i}{\sqrt{A_i^2 + Ce^{2\lambda_i f_i}} + A_i}$$

$$\sqrt{A_i^2 + Ce^{2\lambda_i f_i}} = \frac{A_i + A_i e^{2\lambda_i A_i(t-t_1)}}{1 - e^{2\lambda_i A_i(t-t_1)}}$$

$$Ce^{2\lambda_i f_i} = \frac{4A_i^2 e^{2\lambda_i A_i(t-t_1)}}{1 + e^{4\lambda_i A_i(t-t_1)} - 2e^{2\lambda_i A_i(t-t_1)}}$$

$$= \frac{A_i^2}{\sinh^2\left[\lambda_i A_i(t-t_1)\right]}$$

$$f_i = \frac{1}{\lambda_i} \ln \left\{ \frac{A_i \prod_{k=1}^{n} \left( \frac{m_i - 1}{m_k - 1} \right)^{\frac{m_k}{2}}}{\sinh \left[ \lambda_i A_i (t - t_1) \right]} \right\}.$$  (AB-11)

Now, we get the solution of $f_i$ and then we substitute it into the equation (AB-3) to get the $B_i$ and is given as follow:

$$B_i = \frac{-d\alpha_0 t}{m - 1} + \sqrt{\frac{m_i - 1}{m - 1}} \frac{1}{\lambda_i} \ln \left\{ \frac{A_i \prod_{k=1}^{n} \left( \frac{m_i - 1}{m_k - 1} \right)^{\frac{m_k}{2}}}{\sinh \left[ \lambda_i A_i (t - t_1) \right]} \right\}.$$  (AB-12)

(ii) $k = 0$

Equation (AB-9) becomes

$$\frac{df_i}{A_i} = dt,$$  (AB-13)

and then

$$\int \frac{df_i}{A_i} = \int dt$$

$$f_i = A_i (t - t_1).$$  (AB-14)

Now, we get the solution of $f_i$ and then we substitute it into the equation (AB-3) to get the $B_i$ and is given as follow:

$$B_i = \frac{-d\alpha_0 t}{m - 1} + \sqrt{\frac{m_i - 1}{m - 1}} A_i (t - t_1).$$  (AB-15)

(iii) $k = +1$

Equation (AB-9) becomes

$$\frac{df_i}{\sqrt{A_i^2 - C_i e^{2\lambda_i f_i}}} = dt, \qquad \text{(AB-16)}$$

and then

$$\int \frac{df_i}{\sqrt{A_i^2 - C_i e^{2\lambda_i f_i}}} = \int dt$$

$$\frac{1}{2\lambda_i A_i} \ln\left(\frac{-\sqrt{A_i^2 - Ce^{2\lambda_i f_i}} + A_i}{\sqrt{A_i^2 - Ce^{2\lambda_i f_i}} + A_i}\right) = t - t_1 \qquad (t_1 \text{ is integration constant})$$

$$e^{2\lambda_i A_i (t - t_1)} = \frac{-\sqrt{A_i^2 - Ce^{2\lambda_i f_i}} + A_i}{\sqrt{A_i^2 - Ce^{2\lambda_i f_i}} + A_i}$$

$$\sqrt{A_i^2 - Ce^{2\lambda_i f_i}} = \frac{A_i - A_i e^{2\lambda_i A_i (t - t_1)}}{1 + e^{2\lambda_i A_i (t - t_1)}}$$

$$Ce^{2\lambda_i f_i} = \frac{4 A_i^2 e^{2\lambda_i A_i (t - t_1)}}{1 + e^{4\lambda_i A_i (t - t_1)} + 2 e^{2\lambda_i A_i (t - t_1)}}$$

$$= \frac{A_i^2}{\cosh^2\left[\lambda_i A_i (t - t_1)\right]}$$

$$f_i = \frac{1}{\lambda_i} \ln\left\{\frac{A_i \prod_{k=1}^{n}\left(\frac{m_i - 1}{m_k - 1}\right)^{\frac{m_k}{2}}}{\cosh\left[\lambda_i A_i (t - t_1)\right]}\right\}. \qquad \text{(AB-17)}$$

Now, we get the solution of $f_i$ and then we substitute it into the equation (AB-3) to get the $B_i$ and is given as follow:

$$\boxed{B_i = \frac{-d\alpha_0 t}{m-1} + \sqrt{\frac{m_i-1}{m-1}} \frac{1}{\lambda_i} \ln\left\{\frac{A_i \prod_{k=1}^{n}\left(\frac{m_i-1}{m_k-1}\right)^{\frac{m_k}{2}}}{\cosh\left[\lambda_i A_i (t-t_1)\right]}\right\}} . \qquad (AB\text{-}18)$$

∴ Now, we get the 3 different $B_i$ for 3 different curvatures of internal spaces in equations (AB-12), (AB-15) and (AB-18).

# Appendix C

## Some Detailed calculations of section 4.2.2

(1) Derivation of (4.2.2-2):

From equation (4.2-17),

$$\ddot{B}_3 + k_3(m_3-1)e^{2(m_3-1)B_3+2m_0B_0+2m_1B_1+2m_2B_2} = 0$$

$$\frac{-\ddot{B}_3}{k_3(m_3-1)} = e^{2(m_3-1)B_3+2m_0B_0+2m_1B_1+2m_2B_2}$$

$$\ln\left[\frac{-\ddot{B}_3}{k_3(m_3-1)}\right] = 2(m_3-1)B_3 + 2m_0B_0 + 2m_1B_1 + 2m_2B_2$$

$$B_0 = \frac{1}{2m_0}\ln\left[\frac{-\ddot{B}_3}{k_3(m_3-1)}\right] - \frac{(m_3-1)B_3 + m_1B_1 + m_2B_2}{m_0} \qquad \text{(AC-1)}$$

(2) Derivation of (4.2.2-3) can be seen from (AB-4).

(3) Derivation of (4.2.2-4):

From equation (AB-4),

$$\ddot{B}_1 e^{2B_1} k_3(m_3-1) = k_1(m_1-1)\ddot{B}_3 e^{2B_3}$$
$$-\ddot{B}_1 e^{2B_1}(m_3-1) = 0$$
$$\ddot{B}_1 = 0$$
$$\therefore \ddot{a} = 0 \qquad \text{(AC-2)}$$

and

$$\ddot{B}_2 e^{2B_2} k_3 (m_3 - 1) = k_2 (m_2 - 1) \ddot{B}_3 e^{2B_3}$$
$$-\ddot{B}_2 e^{2B_2} (m_3 - 1) = 0$$
$$\ddot{B}_2 = 0$$
$$\therefore \ddot{b} = 0 \tag{AC-3}$$

(4) Derivation of (4.2.2-5):

From equation (AC-1),

$$\lambda_0 t = \frac{1}{6} \ln\left(\frac{\ddot{c}}{m_3 - 1}\right) - \frac{2(m_3 - 1)\left(c - \frac{3\lambda_0 t}{m-1}\right) + 2m_1\left(a - \frac{3\lambda_0 t}{m-1}\right) + 2m_2\left(b - \frac{3\lambda_0 t}{m-1}\right)}{6}$$

$$6\lambda_0 t = \ln\left(\frac{\ddot{c}}{m_3 - 1}\right) + \frac{6\lambda_0 t}{m-1}(m_1 + m_2 + m_3 - 1) - 2c(m_3 - 1) - 2m_1 a - 2m_2 b$$

$$\ln\left(\frac{\ddot{c}}{m_3 - 1}\right) = 2c(m_3 - 1) + 2m_1 a + 2m_2 b$$

$$\frac{\ddot{c}}{m_3 - 1} = e^{2c(m_3 - 1) + 2m_1 a + 2m_2 b}$$

$$\ddot{c} = (m_3 - 1) e^{2c(m_3 - 1) + 2m_1 a + 2m_2 b} \tag{AC-4}$$

(5) Derivation of (4.2.2-6):

From equation (4.2-16),

$$-\underbrace{\sum_{i=0}^{n} m_i \ddot{B}_i}_{part1} + \underbrace{\sum_{i=0}^{n} m_i (m_i - 1) \dot{B}_i^2}_{part2} + \underbrace{\sum_{j \neq i}^{n} m_i m_j \dot{B}_i \dot{B}_j}_{part3} = 0. \tag{AC-5}$$

I divide it into 3 parts and will evaluate one by one.

Part 1:

$$-\sum_{i=0}^{n} m_i \ddot{B}_i$$

$$= -3\ddot{B}_0 - m_1 \ddot{B}_1 - m_2 \ddot{B}_2 - m_3 \ddot{B}_3$$

$$= -m_3 \ddot{c} \tag{AC-6}$$

Part 2:

$$\sum_{i=0}^{n} m_i (m_i - 1) \dot{B}_i^2$$

$$= m_0 (m_0 - 1) \dot{B}_0^2 + m_1 (m_1 - 1) \dot{B}_1^2 + m_2 (m_2 - 1) \dot{B}_2^2 + m_3 (m_3 - 1) \dot{B}_3^2$$

$$= 6\lambda_0^2 + m_1 (m_1 - 1)\left(\dot{a} - \frac{3\lambda_0}{m-1}\right)^2 + m_2 (m_2 - 1)\left(\dot{b} - \frac{3\lambda_0}{m-1}\right)^2 + m_3 (m_3 - 1)\left(\dot{c} - \frac{3\lambda_0}{m-1}\right)^2$$

$$= 6\lambda_0^2 + m_1 (m_1 - 1)\left(\dot{a}^2 + \frac{9\lambda_0^2}{(m-1)^2} - \frac{6\lambda_0 \dot{a}}{m-1}\right) + m_2 (m_2 - 1)\left(\dot{b}^2 + \frac{9\lambda_0^2}{(m-1)^2} - \frac{6\lambda_0 \dot{b}}{m-1}\right)$$

$$+ m_3 (m_3 - 1)\left(\dot{c}^2 + \frac{9\lambda_0^2}{(m-1)^2} - \frac{6\lambda_0 \dot{c}}{m-1}\right)$$

$$= 6\lambda_0^2 + m_1 (m_1 - 1)\dot{a}^2 + \frac{9\lambda_0^2 m_1 (m_1 - 1)}{(m-1)^2} - \frac{6\lambda_0 \dot{a} m_1 (m_1 - 1)}{m-1} + m_2 (m_2 - 1)\dot{b}^2 + \frac{9\lambda_0^2 m_2 (m_2 - 1)}{(m-1)^2}$$

$$- \frac{6\lambda_0 \dot{b} m_2 (m_2 - 1)}{m-1} + m_3 (m_3 - 1)\dot{c}^2 + \frac{9\lambda_0^2 m_3 (m_3 - 1)}{(m-1)^2} - \frac{6\lambda_0 \dot{c} m_3 (m_3 - 1)}{m-1}$$

$$\tag{AC-7}$$

Part 3:

$$\sum_{j \neq i}^{n} m_i m_j \dot{B}_i \dot{B}_j$$

$$= m_0 m_1 \dot{B}_0 \dot{B}_1 + m_0 m_2 \dot{B}_0 \dot{B}_2 + m_0 m_3 \dot{B}_0 \dot{B}_3 + m_1 m_0 \dot{B}_1 \dot{B}_0 + m_1 m_2 \dot{B}_1 \dot{B}_2 + m_1 m_3 \dot{B}_1 \dot{B}_3$$
$$+ m_2 m_0 \dot{B}_2 \dot{B}_0 + m_2 m_1 \dot{B}_2 \dot{B}_1 + m_2 m_3 \dot{B}_2 \dot{B}_3 + m_3 m_0 \dot{B}_3 \dot{B}_0 + m_3 m_1 \dot{B}_3 \dot{B}_1 + m_3 m_2 \dot{B}_3 \dot{B}_2$$

$$= 2 m_0 m_1 \dot{B}_0 \dot{B}_1 + 2 m_0 m_2 \dot{B}_0 \dot{B}_2 + 2 m_0 m_3 \dot{B}_0 \dot{B}_3 + 2 m_1 m_2 \dot{B}_1 \dot{B}_2 + 2 m_1 m_3 \dot{B}_1 \dot{B}_3 + 2 m_2 m_3 \dot{B}_2 \dot{B}_3$$

$$= 6 m_1 \lambda_0 \left( \dot{a} - \frac{3\lambda_0}{m-1} \right) + 6 m_2 \lambda_0 \left( \dot{b} - \frac{3\lambda_0}{m-1} \right) + 6 m_3 \lambda_0 \left( \dot{c} - \frac{3\lambda_0}{m-1} \right)$$
$$+ 2 m_1 m_2 \left( \dot{a} - \frac{3\lambda_0}{m-1} \right) \left( \dot{b} - \frac{3\lambda_0}{m-1} \right) + 2 m_1 m_3 \left( \dot{a} - \frac{3\lambda_0}{m-1} \right) \left( \dot{c} - \frac{3\lambda_0}{m-1} \right)$$
$$+ 2 m_2 m_3 \left( \dot{b} - \frac{3\lambda_0}{m-1} \right) \left( \dot{c} - \frac{3\lambda_0}{m-1} \right)$$

$$= 6 m_1 \lambda_0 \dot{a} - \frac{18 m_1 \lambda_0^2}{m-1} + 6 m_2 \lambda_0 \dot{b} - \frac{18 m_2 \lambda_0^2}{m-1} + 6 m_3 \lambda_0 \dot{c} - \frac{18 m_3 \lambda_0^2}{m-1}$$
$$+ 2 m_1 m_2 \left[ \dot{a}\dot{b} - \frac{3\dot{a}\lambda_0}{m-1} - \frac{3\dot{b}\lambda_0}{m-1} + \frac{9\lambda_0^2}{(m-1)^2} \right] + 2 m_1 m_3 \left[ \dot{a}\dot{c} - \frac{3\dot{a}\lambda_0}{m-1} - \frac{3\dot{c}\lambda_0}{m-1} + \frac{9\lambda_0^2}{(m-1)^2} \right]$$
$$+ 2 m_2 m_3 \left[ \dot{b}\dot{c} - \frac{3\dot{b}\lambda_0}{m-1} - \frac{3\dot{c}\lambda_0}{m-1} + \frac{9\lambda_0^2}{(m-1)^2} \right]$$

$$= 6 m_1 \lambda_0 \dot{a} - \frac{18 m_1 \lambda_0^2}{m-1} + 6 m_2 \lambda_0 \dot{b} - \frac{18 m_2 \lambda_0^2}{m-1} + 6 m_3 \lambda_0 \dot{c} - \frac{18 m_3 \lambda_0^2}{m-1} + 2 m_1 m_2 \dot{a}\dot{b} - \frac{6 m_1 m_2 \dot{a} \lambda_0}{m-1}$$
$$- \frac{6 m_1 m_2 \dot{b} \lambda_0}{m-1} + \frac{18 m_1 m_2 \lambda_0^2}{(m-1)^2} + 2 m_1 m_3 \dot{a}\dot{c} - \frac{6 m_1 m_3 \dot{a} \lambda_0}{m-1} - \frac{6 m_1 m_3 \dot{c} \lambda_0}{m-1} + \frac{18 m_1 m_3 \lambda_0^2}{(m-1)^2} + 2 m_2 m_3 \dot{b}\dot{c}$$
$$- \frac{6 m_2 m_3 \dot{b} \lambda_0}{m-1} - \frac{6 m_2 m_3 \dot{c} \lambda_0}{m-1} + \frac{18 m_2 m_3 \lambda_0^2}{(m-1)^2}$$

(AC-8)

So, now

$$-\sum_{i=0}^{n} m_i \ddot{B}_i + \sum_{i=0}^{n} m_i(m_i-1)\dot{B}_i^2 + \sum_{j\neq i}^{n} m_i m_j \dot{B}_i \dot{B}_j = 0$$
$$\underbrace{\phantom{-\sum_{i=0}^{n} m_i \ddot{B}_i}}_{part\,1} \underbrace{\phantom{\sum_{i=0}^{n} m_i(m_i-1)\dot{B}_i^2}}_{part\,2} \underbrace{\phantom{\sum_{j\neq i}^{n} m_i m_j \dot{B}_i \dot{B}_j}}_{part\,3}$$

$$-m_3\ddot{c} + 6\lambda_0^2 + m_1(m_1-1)\dot{a}^2 + \frac{9\lambda_0^2 m_1(m_1-1)}{(m-1)^2} - \frac{6\lambda_0 \dot{a} m_1(m_1-1)}{m-1} + m_2(m_2-1)\dot{b}^2$$

$$+\frac{9\lambda_0^2 m_2(m_2-1)}{(m-1)^2} - \frac{6\lambda_0 \dot{b} m_2(m_2-1)}{m-1} + m_3(m_3-1)\dot{c}^2 + \frac{9\lambda_0^2 m_3(m_3-1)}{(m-1)^2} - \frac{6\lambda_0 \dot{c} m_3(m_3-1)}{m-1}$$

$$+6m_1\lambda_0\dot{a} - \frac{18m_1\lambda_0^2}{m-1} + 6m_2\lambda_0\dot{b} - \frac{18m_2\lambda_0^2}{m-1} + 6m_3\lambda_0\dot{c} - \frac{18m_3\lambda_0^2}{m-1} + 2m_1m_2\dot{a}\dot{b} - \frac{6m_1m_2\dot{a}\lambda_0}{m-1}$$

$$-\frac{6m_1m_2\dot{b}\lambda_0}{m-1} + \frac{18m_1m_2\lambda_0^2}{(m-1)^2} + 2m_1m_3\dot{a}\dot{c} - \frac{6m_1m_3\dot{a}\lambda_0}{m-1} - \frac{6m_1m_3\dot{c}\lambda_0}{m-1} + \frac{18m_1m_3\lambda_0^2}{(m-1)^2} + 2m_2m_3\dot{b}\dot{c}$$

$$-\frac{6m_2m_3\dot{b}\lambda_0}{m-1} - \frac{6m_2m_3\dot{c}\lambda_0}{m-1} + \frac{18m_2m_3\lambda_0^2}{(m-1)^2} = 0$$

$$-m_3\ddot{c} + m_1(m_1-1)\dot{a}^2 + m_2(m_2-1)\dot{b}^2 + m_3(m_3-1)\dot{c}^2 + 2m_1m_2\dot{a}\dot{b} + 2m_1m_3\dot{a}\dot{c}$$
$$+ 2m_2m_3\dot{b}\dot{c} = \frac{3\lambda_0^2(m+2)}{(m-1)} \qquad (AC-9)$$

(6) Derivation of the first equation of (4.2.2-7):

From equation (AC-2),

$$\ddot{a} = 0$$

$\therefore a = \alpha_0 t$ (Set the integration constant is zero.) \hfill (AC-10)

(7) Derivation of the second equation of (4.2.2-7):

From equation (AC-3),

$$\ddot{b} = 0$$

$\therefore b = \beta_0 t$ (Set the integration constant is zero.) \hfill (AC-11)

(8) Derivation of the third equation of (4.2.2-7):

We define

$$c = -\frac{m_1 \alpha_0 t}{m_3 - 1} - \frac{m_2 \beta_0 t}{m_3 - 1} + \frac{f}{m_3 - 1}. \tag{AC-12}$$

From equation (AC-4),

$$\ddot{c} = (m_3 - 1) e^{2c(m_3 - 1) + 2m_1 a + 2m_2 b}$$

$$= (m_3 - 1) e^{2(m_3 - 1)\left(-\frac{m_1 \alpha_0 t}{m_3 - 1} - \frac{m_2 \beta_0 t}{m_3 - 1} + \frac{f}{m_3 - 1}\right) + 2m_1 \alpha_0 t + 2m_2 \beta_0 t}$$

$$= (m_3 - 1) e^{2f} \tag{AC-13}$$

Now, we substitute this into the equation (AC-9). I get

$$-m_3 (m_3 - 1) e^{2f} + m_1 (m_1 - 1) \alpha_0^2 + m_2 (m_2 - 1) \beta_0^2 + m_3 (m_3 - 1)\left(-\frac{m_1 \alpha_0}{m_3 - 1} - \frac{m_2 \beta_0}{m_3 - 1} + \frac{\dot{f}}{m_3 - 1}\right)^2$$

$$+ 2m_1 m_2 \alpha_0 \beta_0 + 2m_1 m_3 \alpha_0 \left(-\frac{m_1 \alpha_0}{m_3 - 1} - \frac{m_2 \beta_0}{m_3 - 1} + \frac{\dot{f}}{m_3 - 1}\right) + 2m_2 m_3 \beta_0 \left(-\frac{m_1 \alpha_0}{m_3 - 1} - \frac{m_2 \beta_0}{m_3 - 1} + \frac{\dot{f}}{m_3 - 1}\right)$$

$$= \frac{3\lambda_0^2 (m+2)}{(m-1)}$$

$$-m_3 (m_3 - 1) e^{2f} + m_1 (m_1 - 1) \alpha_0^2 + m_2 (m_2 - 1) \beta_0^2 + \frac{m_1^2 m_3 \alpha_0^2}{m_3 - 1} + \frac{m_1 m_2 m_3 \alpha_0 \beta_0}{m_3 - 1} - \frac{m_1 m_3 \alpha_0 \dot{f}}{m_3 - 1}$$

$$+ \frac{m_1 m_2 m_3 \alpha_0 \beta_0}{m_3 - 1} + \frac{m_2^2 m_3 \beta_0^2}{m_3 - 1} - \frac{m_2 m_3 \beta_0 \dot{f}}{m_3 - 1} - \frac{m_1 m_3 \alpha_0 \dot{f}}{m_3 - 1} - \frac{m_2 m_3 \beta_0 \dot{f}}{m_3 - 1} + \frac{m_3 \dot{f}^2}{m_3 - 1} + 2m_1 m_2 \alpha_0 \beta_0$$

$$- \frac{2m_1^2 m_3 \alpha_0^2}{m_3 - 1} - \frac{2m_1 m_2 m_3 \alpha_0 \beta_0}{m_3 - 1} + \frac{2m_1 m_3 \alpha_0 \dot{f}}{m_3 - 1} - \frac{2m_1 m_2 m_3 \alpha_0 \beta_0}{m_3 - 1} - \frac{2m_2^2 m_3 \beta_0^2}{m_3 - 1} + \frac{2m_2 m_3 \beta_0 \dot{f}}{m_3 - 1}$$

$$= \frac{3\lambda_0^2 (m+2)}{(m-1)}$$

$$-e^{2f} + \frac{\dot{f}^2}{(m_3-1)^2} = \frac{3\lambda_0^2(m+2)}{m_3(m-1)(m_3-1)} + \frac{m_1\alpha_0^2(m_1+m_3-1)}{m_3(m_3-1)^2} + \frac{m_2\beta_0^2(m_2+m_3-1)}{m_3(m_3-1)^2}$$
$$+ \frac{2m_1m_2\alpha_0\beta_0}{m_3(m_3-1)^2}$$

$$-e^{2f} + \frac{\dot{f}^2}{(m_3-1)^2} = \sigma^2$$

where $\sigma^2 = \dfrac{3\lambda_0^2(m+2)}{m_3(m-1)(m_3-1)} + \dfrac{m_1\alpha_0^2(m_1+m_3-1)}{m_3(m_3-1)^2} + \dfrac{m_2\beta_0^2(m_2+m_3-1)}{m_3(m_3-1)^2} + \dfrac{2m_1m_2\alpha_0\beta_0}{m_3(m_3-1)^2}$

(AC-14)

$$\left(\frac{df}{dt}\right)^2 = (\sigma^2 + e^{2f})(m_3-1)^2$$

$$\frac{df}{dt} = (m_3-1)\sqrt{\sigma^2 + e^{2f}}$$

$$\int \frac{df}{(m_3-1)\sqrt{\sigma^2 + e^{2f}}} = \int dt$$

$$\frac{1}{2\sigma(m_3-1)} \ln \frac{\sqrt{\sigma^2+e^{2f}} - \sigma}{\sqrt{\sigma^2+e^{2f}} + \sigma} = t \text{ (Set the integration constant is zero.)}$$

$$\ln \frac{\sqrt{\sigma^2+e^{2f}} - \sigma}{\sqrt{\sigma^2+e^{2f}} + \sigma} = 2\sigma(m_3-1)t$$

$$e^{2\sigma t(m_3-1)} = \frac{\sqrt{\sigma^2+e^{2f}} - \sigma}{\sqrt{\sigma^2+e^{2f}} + \sigma}$$

$$\left(\sqrt{\sigma^2+e^{2f}} + \sigma\right)e^{2\sigma t(m_3-1)} = \sqrt{\sigma^2+e^{2f}} - \sigma$$

$$e^{2\sigma t(m_3-1)}\sqrt{\sigma^2+e^{2f}} + \sigma e^{2\sigma t(m_3-1)} = \sqrt{\sigma^2+e^{2f}} - \sigma$$

$$\sqrt{\sigma^2+e^{2f}}\left(e^{2\sigma t(m_3-1)} - 1\right) = -\sigma - \sigma e^{2\sigma t(m_3-1)}$$

$$\sqrt{\sigma^2 + e^{2f}} = \frac{-\sigma - \sigma e^{2\sigma t(m_3-1)}}{e^{2\sigma t(m_3-1)} - 1}$$

$$\sigma^2 + e^{2f} = \frac{\left(-\sigma - \sigma e^{2\sigma t(m_3-1)}\right)\left(-\sigma - \sigma e^{2\sigma t(m_3-1)}\right)}{\left(e^{2\sigma t(m_3-1)} - 1\right)\left(e^{2\sigma t(m_3-1)} - 1\right)}$$

$$= \frac{\sigma^2 + 2\sigma^2 e^{2\sigma t(m_3-1)} + \sigma^2 e^{4\sigma t(m_3-1)}}{e^{4\sigma t(m_3-1)} - 2e^{2\sigma t(m_3-1)} + 1}$$

$$e^{2f} = \frac{\sigma^2 + 2\sigma^2 e^{2\sigma t(m_3-1)} + \sigma^2 e^{4\sigma t(m_3-1)}}{e^{4\sigma t(m_3-1)} - 2e^{2\sigma t(m_3-1)} + 1} - \frac{\sigma^2\left(e^{4\sigma t(m_3-1)} - 2e^{2\sigma t(m_3-1)} + 1\right)}{\left(e^{4\sigma t(m_3-1)} - 2e^{2\sigma t(m_3-1)} + 1\right)}$$

$$= \frac{4\sigma^2 e^{2\sigma t(m_3-1)}}{e^{4\sigma t(m_3-1)} - 2e^{2\sigma t(m_3-1)} + 1}$$

$$= \frac{\sigma^2}{\dfrac{1}{4}e^{2\sigma t(m_3-1)} + \dfrac{1}{4}e^{-2\sigma t(m_3-1)} - \dfrac{2}{4}}$$

$$= \frac{\sigma^2}{\left(\dfrac{1}{2}e^{\sigma t(m_3-1)} - \dfrac{1}{2}e^{-\sigma t(m_3-1)}\right)^2}$$

$$= \frac{\sigma^2}{\sinh^2\left[\sigma t(m_3-1)\right]}$$

$$2f = \ln\left(\frac{\sigma^2}{\sinh^2\left[\sigma t(m_3-1)\right]}\right)$$

$$f = \ln\left(\frac{\sigma}{\sinh\left[\sigma t(m_3-1)\right]}\right) \tag{AC-15}$$

Now, we get the solution of $f$ and then we substitute it into the equation (AC-12) to get the $c$ and is given as follow:

$$c = -\frac{m_1 \alpha_0 t}{m_3 - 1} - \frac{m_2 \beta_0 t}{m_3 - 1} + \frac{\ln\left(\dfrac{\sigma}{\sinh[\sigma t(m_3 - 1)]}\right)}{m_3 - 1}. \tag{AC-16}$$

(9) Derivation of (4.2.2-8):

From equation (4.2.1-5),

$$ds^2_{E,d+1} = e^{\frac{2}{d-1}\sum_{i=1}^{n} m_i B_i}\left(-e^{2\sum_{j=0}^{n} m_j B_j} dt^2 + e^{2B_0} ds^2_d\right)$$

$$= -e^{\frac{2}{d-1}\sum_{i=1}^{n} m_i B_i + 2\sum_{j=0}^{n} m_j B_j} dt^2 + e^{\frac{2}{d-1}\sum_{i=1}^{n} m_i B_i + 2B_0} ds^2_d$$

$$= -e^{\frac{2}{d-1}\left(a - \frac{3\lambda_0 t}{m-1} + 3b - \frac{9\lambda_0 t}{m-1} + 3c - \frac{9\lambda_0 t}{m-1}\right) + 2\left(d\lambda_0 t + a - \frac{3\lambda_0 t}{m-1} + 3b - \frac{9\lambda_0 t}{m-1} + 3c - \frac{9\lambda_0 t}{m-1}\right)} dt^2$$

$$+ e^{\frac{2}{d-1}\left(a - \frac{3\lambda_0 t}{m-1} + 3b - \frac{9\lambda_0 t}{m-1} + 3c - \frac{9\lambda_0 t}{m-1}\right) + 2\lambda_0 t} ds^2_d$$

$$= -e^{\frac{2}{d-1}\left(a - \frac{3\lambda_0 t}{m-1} + 3b - \frac{9\lambda_0 t}{m-1} + 3c - \frac{9\lambda_0 t}{m-1}\right) + 2\left(a - \frac{3\lambda_0 t}{m-1} + 3b - \frac{9\lambda_0 t}{m-1} + 3c - \frac{9\lambda_0 t}{m-1}\right) + 2d\lambda_0 t} dt^2$$

$$+ e^{\frac{2}{d-1}\left(a - \frac{3\lambda_0 t}{m-1} + 3b - \frac{9\lambda_0 t}{m-1} + 3c - \frac{9\lambda_0 t}{m-1}\right) + 2\lambda_0 t} ds^2_d$$

$$= -e^{\left[\frac{2}{d-1} + \frac{2(d-1)}{(d-1)}\right]\left(a - \frac{3\lambda_0 t}{m-1} + 3b - \frac{9\lambda_0 t}{m-1} + 3c - \frac{9\lambda_0 t}{m-1}\right) + 2d\lambda_0 t} dt^2 + e^{\frac{2}{d-1}\left(a - \frac{3\lambda_0 t}{m-1} + 3b - \frac{9\lambda_0 t}{m-1} + 3c - \frac{9\lambda_0 t}{m-1}\right) + 2\lambda_0 t} ds^2_d$$

$$= -e^{\frac{2d}{d-1}\left(a - \frac{3\lambda_0 t}{m-1} + 3b - \frac{9\lambda_0 t}{m-1} + 3c - \frac{9\lambda_0 t}{m-1}\right) + 2d\lambda_0 t} dt^2 + e^{\frac{2}{d-1}\left(a - \frac{3\lambda_0 t}{m-1} + 3b - \frac{9\lambda_0 t}{m-1} + 3c - \frac{9\lambda_0 t}{m-1}\right) + 2\lambda_0 t} ds^2_d$$

$$= -S^{2d}(t) dt^2 + S^2(t) ds^2_{R,d} \tag{AC-17}$$

where $S = e^{\frac{1}{d-1}\left(a - \frac{3\lambda_0 t}{m-1} + 3b - \frac{9\lambda_0 t}{m-1} + 3c - \frac{9\lambda_0 t}{m-1}\right) + \lambda_0 t}$

(10) Derivation of (4.2.2-13):

For the case $c_1 = \frac{1}{2}, c_2 = 2, m_1 = 1, m_2 = 3, m_3 = 3$, and from (AC-14),

$$\sigma^2 = \frac{3\lambda_0^2(m+2)}{m_3(m-1)(m_3-1)} + \frac{m_1\alpha_0^2(m_1+m_3-1)}{m_3(m_3-1)^2} + \frac{m_2\beta_0^2(m_2+m_3-1)}{m_3(m_3-1)^2} + \frac{2m_1m_2\alpha_0\beta_0}{m_3(m_3-1)^2}$$

$$= \frac{3\lambda_0^2(7+2)}{3(7-1)(3-1)} + \frac{\left(\frac{1}{2}\lambda_0\right)^2(1+3-1)}{3(3-1)^2} + \frac{3(2\lambda_0)^2(3+3-1)}{3(3-1)^2} + \frac{2(3)\left(\frac{1}{2}\lambda_0\right)(2\lambda_0)}{3(3-1)^2}$$

$$= \frac{\lambda_0^2 \, 101}{16}$$

$$\sigma = \frac{\lambda_0\sqrt{101}}{4} \tag{AC-18}$$

(11) Derivation of (4.2.2-14):

From equation (AC-17),

$$S = e^{\frac{1}{d-1}\left(a - \frac{3\lambda_0 t}{m-1} + 3b - \frac{9\lambda_0 t}{m-1} + 3c - \frac{9\lambda_0 t}{m-1}\right) + \lambda_0 t}$$

$$= e^{\frac{1}{2}\left[\frac{1}{2}\lambda_0 t - \frac{3\lambda_0 t}{6} + 3(2\lambda_0 t) - \frac{9\lambda_0 t}{6} + 3c - \frac{9\lambda_0 t}{6}\right] + \lambda_0 t}$$

$$= e^{\frac{5\lambda_0 t}{2} + \frac{3c}{2}} \tag{AC-19}$$

(12) Derivation of (4.2.2-15):

From equation (AC-19),

$$S = e^{\frac{5\lambda_0 t}{2} + \frac{3c}{2}}$$

$$= e^{\frac{5\lambda_0 t}{2} + \frac{3}{2}\left[-\frac{\alpha_0 t}{2} - \frac{3\beta_0 t}{2} + \frac{1}{2}\ln\left(\frac{\sigma}{\sinh 2t\sigma}\right)\right]}$$

$$= e^{\frac{5\lambda_0 t}{2} + \frac{3}{2}\left[-\frac{\left(\frac{1}{2}\lambda_0\right)t}{2} - \frac{3(2\lambda_0)t}{2} + \frac{1}{2}\ln\left(\frac{\sigma}{\sinh 2t\sigma}\right)\right]}$$

$$= e^{-\frac{19\lambda_0 t}{8} + \frac{3}{4}\ln\left(\frac{\sigma}{\sinh 2t\sigma}\right)} \tag{AC-20}$$

By differentiating the (AC-20) with respect to t, we get

$$\frac{dS}{dt} = e^{-\frac{19\lambda_0 t}{8} + \frac{3}{4}\ln\left(\frac{\sigma}{\sinh 2t\sigma}\right)} \times \left[-\frac{19\lambda_0}{8} - \frac{3\sigma}{2}(\coth 2t\sigma)\right], \tag{AC-21}$$

and also from definition,

$$\frac{dt}{d\tau} = \frac{1}{S^d}$$

$$= \left[e^{-\frac{19\lambda_0 t}{8} + \frac{3}{4}\ln\left(\frac{\sigma}{\sinh 2t\sigma}\right)}\right]^{-3}$$

$$= e^{\frac{57\lambda_0 t}{8} - \frac{9}{4}\ln\left(\frac{\sigma}{\sinh 2t\sigma}\right)}. \tag{AC-22}$$

Then,

$$\frac{dS}{dt}\frac{dt}{d\tau} = e^{-\frac{19\lambda_0 t}{8} + \frac{3}{4}\ln\left(\frac{\sigma}{\sinh 2t\sigma}\right)} \times \left[-\frac{19\lambda_0}{8} - \frac{3\sigma}{2}(\coth 2t\sigma)\right] \times e^{\frac{57\lambda_0 t}{8} - \frac{9}{4}\ln\left(\frac{\sigma}{\sinh 2t\sigma}\right)}$$

$$= e^{\frac{38\lambda_0 t}{8} - \frac{3}{2}\ln\left(\frac{\sigma}{\sinh 2t\sigma}\right)} \times \left[-\frac{19\lambda_0}{8} - \frac{3\sigma}{2}(\coth 2t\sigma)\right]$$

Therefore, from (4.2.2-10),

$$\frac{d\left(\frac{dS}{dt}\frac{dt}{d\tau}\right)}{d\tau} > 0$$

$$\left[\frac{38\lambda_0}{8} + 3\sigma(\coth 2t\sigma)\right] \times \left[-\frac{19\lambda_0}{8} - \frac{3\sigma}{2}(\coth 2t\sigma)\right] + \left[3\sigma^2\left(\csc h^2 2t\sigma\right)\right] > 0$$

$$\left[\frac{19\lambda_0}{8}+\frac{3\sigma}{2}(\coth 2t\sigma)\right]\times(-1)\left[\frac{19\lambda_0}{8}+\frac{3\sigma}{2}(\coth 2t\sigma)\right]+\frac{3}{2}\sigma^2\left(\csc h^2 2t\sigma\right)>0$$

$$-\left[\frac{19\lambda_0}{8}+\frac{3\sigma}{2}(\coth 2t\sigma)\right]^2>-\frac{3}{2}\sigma^2\left(\csc h^2 2t\sigma\right)$$

$$\frac{3\sigma^2}{2\left(\sinh^2 2t\sigma\right)}>\left[\frac{19\lambda_0}{8}+\frac{3\sigma}{2}(\coth 2t\sigma)\right]^2 \tag{AC-23}$$

(13) Derivation of (4.2.2-16):

From equation (AC-23),

$$\frac{3\sigma^2}{2\left(\sinh^2 2t\sigma\right)}>\left[\frac{19\lambda_0}{8}+\frac{3\sigma}{2}(\coth 2t\sigma)\right]\times\left[\frac{19\lambda_0}{8}+\frac{3\sigma}{2}(\coth 2t\sigma)\right]$$

$$>\frac{361\lambda_0^2}{64}+\frac{114\lambda_0\sigma\coth 2t\sigma}{16}+\frac{9\sigma^2\coth^2 2t\sigma}{4}$$

$$>\frac{5776\sigma^2}{6464}+\frac{456\sigma^2\coth 2t\sigma}{16\sqrt{101}}+\frac{9\sigma^2\coth^2 2t\sigma}{4}$$

$$\frac{3}{2\left(\sinh^2 2t\sigma\right)}>\frac{5776}{6464}+\frac{456\coth 2t\sigma}{16\sqrt{101}}+\frac{9\coth^2 2t\sigma}{4}$$

$$1212\sqrt{101}>722\sqrt{101}\sinh^2\theta+23028\sinh\theta\cosh\theta+1818\sqrt{101}\cosh^2\theta \qquad (\theta=2t\sigma)$$

$$1212\sqrt{101}>722\sqrt{101}\left(\frac{e^\theta-e^{-\theta}}{2}\right)^2+23028\left(\frac{e^\theta-e^{-\theta}}{2}\right)\left(\frac{e^\theta+e^{-\theta}}{2}\right)+1818\sqrt{101}\left(\frac{e^\theta+e^{-\theta}}{2}\right)^2$$

$$1212\sqrt{101}>722\sqrt{101}\left(\frac{e^{2\theta}+e^{-2\theta}-2}{4}\right)+23028\left(\frac{e^{2\theta}-e^{-2\theta}}{4}\right)+1818\sqrt{101}\left(\frac{e^{2\theta}+e^{-2\theta}+2}{4}\right)$$

$$4848\sqrt{101}>722\sqrt{101}e^{2\theta}+722\sqrt{101}e^{-2\theta}-1444\sqrt{101}+23028e^{2\theta}-23028e^{-2\theta}$$
$$+1818\sqrt{101}e^{2\theta}+1818\sqrt{101}e^{-2\theta}+3636\sqrt{101}$$

$$4848\sqrt{101}e^{2\theta} > 722\sqrt{101}e^{4\theta} + 722\sqrt{101} - 1444\sqrt{101}e^{2\theta} + 23028e^{4\theta} - 23028$$

$$+1818\sqrt{101}e^{4\theta} + 1818\sqrt{101} + 3636\sqrt{101}e^{2\theta}$$

$$2656\sqrt{101}e^{2\theta} - 2540\sqrt{101}e^{4\theta} - 23028e^{4\theta} - 2540\sqrt{101} + 23028 > 0$$

$$-48554.6841x^2 + 26692.4697x - 2498.6841 > 0 \qquad \left(x = e^{2\theta}\right)$$

For the interval of $x$ which is satisfied in this inequality is given as following:

$$\left(\frac{-26692.4697 + \sqrt{26692.4697^2 - 4(-48554.6841)(-2498.6841)}}{2(-48554.6841)}\right)$$

$$< x <$$

$$\left(\frac{-26692.4697 - \sqrt{26692.4697^2 - 4(-48554.6841)(-2498.6841)}}{2(-48554.6841)}\right)$$

$$0.1197 < x < 0.4301$$

$$0.1197 < e^{2\theta} < 0.4301$$

$$0.1197 < e^{2(2t\sigma)} < 0.4301$$

$$\ln 0.1197 < 4t\sigma < \ln 0.4301$$

$$\frac{\ln 0.1197}{4\sigma} < t < \frac{\ln 0.4301}{4\sigma} \tag{AC-24}$$

# BIODATA OF THE AUTHOR

Mr. Ch'ng Han Siong was born in 1979 at Bukit Mertajam. His previous secondary schools were Sekolah Menengah Kampong Kastam, Butterworth and Kolej Komuniti Mertajam, Bukit Mertajam. Later he studied in Universiti Kebangsaan Malaysia. He graduated with a Degree of Nuclear Science (Hons.) in year 2003.